\documentclass{article}
 \pdfoutput=1

\usepackage{arxiv}
\usepackage[utf8]{inputenc} 
\usepackage[T1]{fontenc}    
\usepackage{hyperref}       
\usepackage{url}            
\usepackage{booktabs}       
\usepackage{amsfonts}       
\usepackage{nicefrac}       
\usepackage{doi}            
\usepackage{fancyhdr}       
\usepackage{graphicx}       

\usepackage{float}
\usepackage{multirow}
\usepackage{array}
\usepackage{bigstrut}
\usepackage{longtable}
\usepackage{makecell}
\usepackage{layouts}

\graphicspath{{./figures/}} 

\title{Altered Topological Properties of Functional Brain Network Associated with Alzheimer’s Disease}

\author{ 
\href{https://orcid.org/0000-0003-2754-3649}{\includegraphics[scale=0.06]{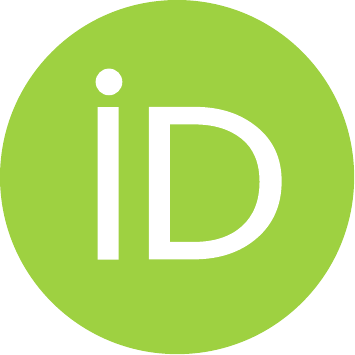}\hspace{1mm} Yongcheng Yao} \\
Department of Imaging and Interventional Radiology \\
The Chinese University of Hong Kong \\
Hong Kong\\
\texttt{yao\_yongchneng@link.cuhk.edu.hk} \\
}

\begin{document}
\maketitle

\begin{abstract}
Functional Magnetic Resonance Imaging (fMRI) is commonly utilized to study human brain activity, including abnormal functional properties related to neurodegenerative diseases. This study aims to investigate the differences in the topological properties of functional brain networks between individuals with Alzheimer's Disease (AD) and normal controls. A total of 590 subjects, consisting of 175 with AD dementia and 415 age-, gender-, and handedness-matched controls, were included. The topological properties of the brain network were quantified using graph-theory-based analyses. The results indicate abnormal network integration and segregation in the AD group. These findings enhance our understanding of AD pathophysiology from a functional brain network structure perspective and may aid in identifying AD biomarkers. Supplementary data to aid in the validation of this research are available at \hyperlink{https://github.com/YongchengYAO/AD-FunctionalBrainNetwork}{https://github.com/YongchengYAO/AD-FunctionalBrainNetwork}. 
\end{abstract}


\section{Introduction}
Alzheimer's Disease (AD) is a chronic neurodegenerative disease prevalent in the elderly population, characterized by cognitive decline, language problems, memory disturbances (especially short-term memory), and disorientation. With disease progression, severe bodily dysfunction and ultimately death can occur. AD is the most common form of dementia, accounting for approximately half of the cases. A rare form of AD is early-onset familial Alzheimer's disease, which is associated with the amyloid precursor protein and presenilin genes. Another form of AD is sporadic AD, affecting over 15 million people worldwide, with its cause primarily unknown. Risk factors for AD include decreased brain size, low education level, low mental ability, head injury, and vascular-disease-related factors \cite{Blennow_2006}. The amyloid hypothesis proposes that extracellular amyloid beta deposits cause AD \cite{hardy1991amyloid}. The tau hypothesis suggests that AD results from tau protein dysfunction, with neurofibrillary tangles formed by tau protein destroying the neuron's transport system \cite{iqbal2005tau}.

Functional magnetic resonance imaging (fMRI) offers a non-invasive approach for diagnosing, evaluating therapeutic interventions, and investigating the mechanisms of AD. In brain imaging, a typical fMRI utilizes the Blood Oxygenation Level Dependent (BOLD) contrast to indirectly reflect brain activity through signal fluctuations. Early studies of brain function primarily relied on task-based fMRI, where fMRI brain activity was acquired during specific functional tasks \cite{pariente2005alzheimer,celone2006alterations}. In 1995, Biswal demonstrated that resting-state fMRI signals could depict spontaneous neuronal activity without the need for external task experiments \cite{biswal1995functional}. An increasing number of studies have utilized resting-state fMRI to investigate brain function and disease-related abnormalities. In recent years, resting-state fMRI has become the most widely used neuroimaging technique in AD-related studies \cite{chen2011classification,wang2006changes,agosta2012resting,binnewijzend2012resting,koch2012diagnostic}.

Graph-theory-based analysis is a prevalent method in brain function studies. To some extent, the graph-based method is also a connectivity analysis, as the interactions among brain regions are described in a network (a graph). In contrast to connectivity analysis, a graph-based analysis investigates the network's topological properties instead of network connections. Topological properties are quantified by various network metrics \cite{rubinov2010complex}.
Previous studies using functional MRI and connectivity analysis have reported disrupted functional connectivity in AD patients. Key regions related to neural degeneration in AD patients include the Hippocampus \cite{wang2006changes,allen2007reduced}, Prefrontal Cortex \cite{wang2007altered}, Parietal Lobe \cite{wang2015differentially}, Precuneus \cite{xia2014differentially}, and Posterior Cingulate Gyrus \cite{zhang2009detection,zhang2010resting,xia2014differentially}. In a systematic review and meta-analysis of 43 studies with 1,363 subjects, regions within the default mode, salience, and limbic networks consistently showed abnormalities in connectivity \cite{badhwar2017resting}. The default mode network has been a focal point of AD studies since changes in functional connectivity in regions of the default mode network have been observed in preclinical AD (from very mild \cite{sorg2007selective} to moderate AD \cite{greicius2004default}). These findings are even more compelling for subjects with severe AD. It has been reported that the map of decreased functional connectivity is similar to the amyloid deposition map and the tau-protein deposition map \cite{brier2014network}, indicating that amyloid deposition \cite{bero2011neuronal} and tau protein deposition may play a causative role in the dysfunction of AD patients. Dysfunction in the default mode network is a prominent feature of AD at all stages. In contrast to the disruption of functional connections in the default mode network, increased functional connections in the salience network have also been reported \cite{zhou2010divergent}, which have been validated by subsequent studies \cite{brier2012loss,brier2014network}. Brier \emph{et al.} \cite{brier2012loss} showed that the pattern of aberrant functional connectivity changes with AD progression. At the early stage of AD (with very mild symptoms), increased functional connectivity within the salience network and decreased functional connectivity within the default mode network, executive control network, and sensorimotor network were observed. However, at the mild to moderate stage, decreased functional connections were discovered in all networks. Brier and colleagues \cite{brier2014network} have proposed a model of disease spread to account for changes in functional disruption in AD.

\section{Graph-Theory-Based Analysis}
\label{chapter:graph.theory.analyses}

\subsection{Related Works}
\paragraph{Graph Type}
The analysis of a graph depends on how it is defined. A graph can be defined as weighted or weightless (binary) based on whether a network connection has weight or not. Additionally, a graph can be directional or bidirectional based on its directionality. Therefore, there are four types of graphs: weighted directional graph, weighted bidirectional graph, weightless directional graph, and weightless bidirectional graph. A weightless graph is a special case of a weighted graph in which all connections have the same weight of 1. Similarly, a bidirectional graph is a unique form of a directional graph, where each bidirectional connection can be viewed as two opposite directional connections. In summary, a weighted and directional graph is the most common form, and the other three types of graphs are special cases under additional conditions.

\paragraph{Network Segregation}
The ability of a network to organize nodes into clusters or modules is termed network segregation. Clusters or modules are groups of nodes that are densely intra-connected and sparsely inter-connected. Measures of segregation \cite{sporns2013network} are primarily designed to identify and quantify these structures, such as the size and number of clusters, or other metrics reflecting the properties of connectivity within a cluster. From the perspective of the functional brain network, functional segregation is the ability to organize the brain into distinct functional subnetworks, each of which is responsible for a relatively unique brain function. Therefore, the functional segregation of the human brain is physiologically meaningful and interpretable.

\paragraph{Network Integration}
The degree of interconnectivity among nodes is reflected in network integration. Measures of integration \cite{sporns2013network} generally quantify how easily nodes communicate with one another, with many of these measures based on the concept of "path." In the context of the functional brain network, functional integration can be viewed as the ability to efficiently retrieve and combine information distributed across different functional subnetworks.

\begin{figure}[]
	\centering
	\includegraphics[width=0.9\linewidth] {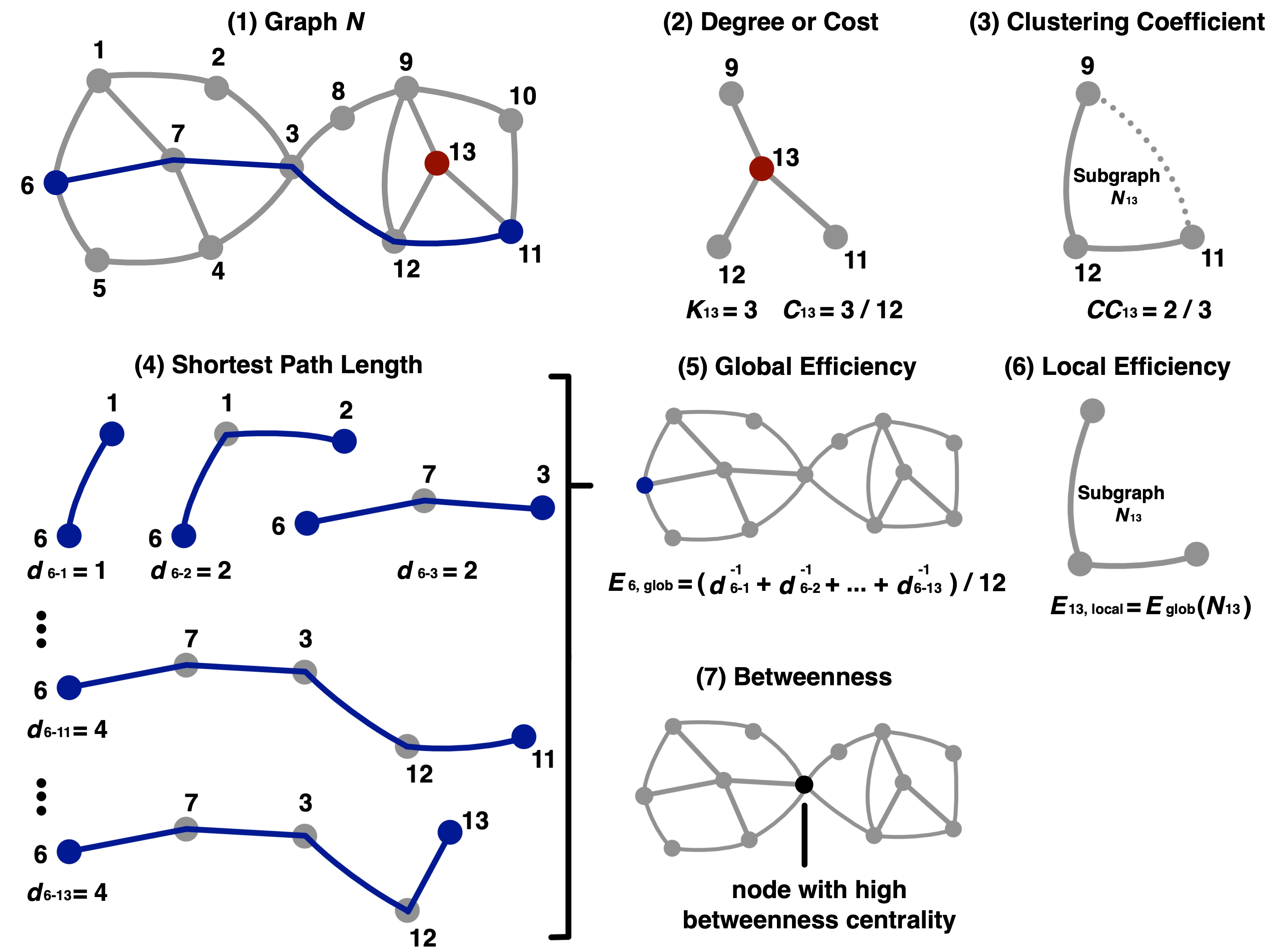}
	\caption[Illustration of Graph Theory Metrics] {{\bf Illustration of Graph Theory Metrics.} (1) A visualization of graph $N$ as interconnected nodes. (2-7) Illustrations of degree / cost, clustering coefficient, shortest path length, global efficiency, local efficiency, and betweenness centrality.}
	\label{fig:illustration.graph.theory}
\end{figure}

\paragraph{Network Metrics}
In this study, we first constructed weightless and bidirectional graphs (brain networks) from MRI data. Second, we quantified the networks' topological properties using graph-theory-based metrics. Finally, significant differences in metrics between groups can depict the topological changes of the networks.
The graph-theory-based metrics used, as shown in Figure \ref{fig:illustration.graph.theory}, are only for weightless bidirectional graphs, where each edge is binary and bidirectional. They are the same as those introduced in an previous review paper \cite{rubinov2010complex}, which includes elaborate explanations and definitions. Additionally, the metric definitions for weighted and directional graphs are also listed (in Table A1 of paper \cite{rubinov2010complex}) and compared with their weightless and bidirectional counterparts. It is noteworthy that metrics of different types of networks cannot be directly compared due to their distinct definitions.

\subsection{Data}
\label{section:graphtheory.subject.data}
\subsubsection{OASIS-3 Dataset}
The subjects involved in our study were obtained from the OASIS-3 public dataset \cite{lamontagne2018oasis}, which is the latest release of the Open Access Series of Imaging Studies (OASIS). OASIS-3 is a large longitudinal dataset that provides the scientific community with open access not only to multi-modal neuroimaging data but also to various clinical data and cognitive assessments. All data in OASIS-3 are available on the OASIS Brains project website (\href{www.oasis-brains.org}{www.oasis-brains.org}). We used the OASIS-3 dataset in this study for several reasons: 
\begin{enumerate}
    \item  Compared to other open-access datasets such as the Alzheimer's Disease Neuroimaging Initiative (ADNI) \cite{weiner2010alzheimer,weiner2013alzheimer,weiner2015impact,weiner2017alzheimer} and the Harvard Aging Brain Study (HABS) \cite{dagley2017harvard}, OASIS-3 is relatively larger in terms of the number of participants (over 1000) and MR sessions (over 2100). It is a retrospective compilation of neuroimaging and clinical data collected across multiple ongoing projects in the past 30 years, involving 609 cognitively normal (normal ageing) adults and 489 subjects at various stages of cognitive decline.
    \item It is easy to explore and download imaging and clinical data. All data in OASIS-3 are hosted by XNAT (the extensible Neuroimaging Archive Toolkit) \cite{marcus2007extensible}, an informatics platform for data management and sharing.
    \item The dataset covers all common modalities of neuroimaging in the field of ageing and cognitive decline. Since the dataset is tailored for research on normal ageing and Alzheimer's disease, it provides open access to over 2100 MR sessions, including T1w, T2w, FLAIR, ASL, SWI, time-of-flight, resting-state BOLD fMRI, and DTI. One of the advantages of OASIS-3 is the degree of data integration, as for most sessions, the majority of modalities are available.
\end{enumerate}

\subsubsection{Inclusion Criteria}
The first step of this study is to select data from a large dataset. Since this is a between-group MRI study, the criteria are related to the parameters of MR images and demographic information. The inclusion criteria are as follows.

\begin{enumerate}
    \item Only data from one session were downloaded for each individual. Although longitudinal data are available, we only chose one BOLD-fMRI and one T1w MRI for each subject. The reason is that this study mainly focuses on investigating the significant differences in topological properties of the functional brain network between the two groups, rather than on longitudinal changes such as disease progression. This rule ensures that the subjects enrolled in this study are independent, which is an assumption of parametric statistical analyses.
    \item The acquisition protocols of BOLD-fMRIs should be the same. Multi-site and multi-scanner MR studies are challenging because of the variabilities introduced by the differences in imaging protocols and scanners. To minimize such variabilities, we exerted restrictions that all fMRI data should be collected from Siemens scanners under the same imaging protocol. However, such a restriction was not applied to T1w MRI data, because structural MRI data were used for brain tissue segmentation to facilitate the removal of BOLD signal artefacts, and the segmentation algorithm used in this study performs well on T1w images with various acquisition parameters.
    \item For each individual, one BOLD-fMRI should be matched with one T1w MRI from the same session. If there is no T1w MRI available in the same session, then the BOLD-fMRI should be discarded, and the subject should be excluded. Since functional MRI data have a low spatial resolution (usually on the scale of 3 mm) while T1-weighted structural MRI data have a relatively high spatial resolution (on the scale of 1 mm), T1w MRI can better delineate the structure of the brain, serving as a spatial reference for BOLD-fMRI. Therefore, T1w MRI and BOLD-fMRI are usually collected in the same session. In other words, the perfect structural reference is the one from the same scanning session.
    \item There should be no significant difference in age, gender, and handedness between the normal control and Alzheimer's Disease group. Significant alterations of brain structure and function were discovered in normal ageing adults. Gender differences and the effect of handedness were also found in functional activation and connectivity. Thus, it is critical to ensure that the two groups are age-, gender-, and handedness-matched.
\end{enumerate}

\subsubsection{MR Image Acquisition Parameters}
Resting-state BOLD MR images were acquired using a single-shot FID EPI sequence on a 3-Tesla scanner (Siemens, TrioTim or Biograph\textunderscore mMR), with the following parameters: TR = 2200 $ms$; TE = 27 $ms$; FA = $90^{\circ}$; slice thickness = 4 $mm$; slice gap = 0 $mm$; number of slices (z) = 36; in-plane resolution = 4 x 4 $mm^{2}$; in-plane matrix size (x, y) = 64 x 64; number of time points = 164.

T1-weighted MR images were acquired using a single-shot TurboFLASH sequence on the same 3-Tesla scanner (Siemens, TrioTim or Biograph\textunderscore mMR), with the following parameters: TR = 2400 $ms$; TE = 3 $ms$; FA = $8^{\circ}$; slice thickness = 1 $mm$; slice gap = 0 $mm$; number of slices (z) = 176; in-plane resolution = 1 x 1 $mm^{2}$; in-plane matrix size (y, z) = 256 x 256.

\subsubsection{MR Image Labeling}
Clinical data, including the Clinical Dementia Rating (CDR) \cite{hughes1982new} \cite{morris1991clinical} and diagnoses, can be downloaded from the "ADRC Clinical Data" field in the OASIS-3 data browser.
The CDR is a 5-point scale used to assess the severity of dementia by characterizing 6 categories of cognitive and functional performance relevant to Alzheimer's Disease and related dementias. These categories are memory, orientation, judgment and problem-solving, community affairs, home and hobbies, and personal care. The rating for each category is obtained through a semi-structured interview, known as the CDR Assessment Protocol. By combining the ratings for each category, a global CDR score can be calculated using the CDR-assignment algorithm \cite{morris1991clinical}. Since clinicians base their diagnoses not only on the global CDR score but also on other clinical tests, we used these professional diagnoses to categorize each BOLD-fMRI into either an AD group or a normal controls (NC) group. Specifically, we used diagnoses from the "dx1" field to identify AD and NC data. A complete list of unique diagnosis remarks and their corresponding labels can be found in Table \ref{tab:diagnoses.labels}.

However, the dates of clinical assessments and MRI scanning are not the same, so it is necessary to link the appropriate clinical diagnoses to each MR image. To accomplish this, we first calculated the relative date of clinical screening from the clinical data identifier. Similarly, the relative date of MRI scanning can be calculated from the name of the MR session. For instance, a clinical data ID "OAS30001\_ClinicalData\_d0339" indicates that the clinical assessments of subject "OAS30001" were performed on the $339^{th}$ day after their first visit. Similarly, an MR session name "OAS30001\_MR\_d0129" indicates that the MRI data of subject "OAS30001" were collected on the $129^{th}$ day after their first visit. Secondly, we linked each MRI data with its nearest clinical results.

\begin{small}
\setlength\LTleft{1.5cm}
\setlength\LTright{1.5cm}
\begin{longtable}[c]{l | c | l | c}
	\caption{Clinical Diagnoses and Labels} \label{tab:diagnoses.labels} \\
\hline \hline 
\multicolumn{1}{c}{\textbf{Diagnosis}} & \multicolumn{1}{c}{\textbf{Label}} & \multicolumn{1}{c}{\textbf{Diagnosis}} & \multicolumn{1}{c}{\textbf{Label}} \\ 
\hline \hline 
\endfirsthead
\multicolumn{4}{l}%
{{\tablename\ \thetable{} -- continued from previous page}} \\
\hline 
\multicolumn{1}{c}{\textbf{Diagnosis}} & \multicolumn{1}{c}{\textbf{Label}} & \multicolumn{1}{c}{\textbf{Diagnosis}} & \multicolumn{1}{c}{\textbf{Label}} \\ 
\hline 
\endhead
\hline \multicolumn{4}{r}{{Continued on next page}} \\ \hline
\endfoot
\hline \hline
\endlastfoot
``AD Dementia'' & AD & ``AD dem w/oth unusual features'' & AD \\
``AD dem Language dysf after'' & AD & ``AD dem w/oth unusual features/demt on'' & AD\\
``AD dem Language dysf prior'' & AD & ``AD dem/FLD prior to AD dem'' & AD\\
``AD dem Language dysf with'' & AD & ``Cognitively normal'' & NC\\
``AD dem cannot be primary'' & AD & ``DAT''   & -\\
``AD dem distrubed social- after'' & AD & ``DAT w/depresss not contribut'' & -\\
``AD dem distrubed social- prior'' & AD & ``DLBD, primary'' & -\\
``AD dem distrubed social- with'' & AD & ``DLBD- primary'' & -\\
``AD dem visuospatial, after'' & AD & ``DLBD- secondary'' & -\\
``AD dem visuospatial- prior'' & AD & ``Dementia/PD- primary'' & -\\
``AD dem visuospatial- with'' & AD & ``Frontotemporal demt. prim'' & -\\
``AD dem w/CVD contribut'' & AD & ``Incipient Non-AD dem'' & -\\
``AD dem w/CVD not contrib'' & AD & ``Incipient demt PTP'' & -\\
``AD dem w/Frontal lobe/demt at onset'' & AD & ``No dementia'' & -\\
``AD dem w/PDI after AD dem contribut'' & AD & ``Non AD dem- Other primary'' & -\\
``AD dem w/PDI after AD dem not contrib'' & AD & ``ProAph w/o dement'' & -\\
``AD dem w/depresss  contribut'' & AD & ``Unc: impair reversible'' & -\\
``AD dem w/depresss  not contribut'' & AD & ``Unc: ques. Impairment'' & -\\
``AD dem w/depresss, not contribut'' & AD & ``Vascular Demt  primary'' & -\\
``AD dem w/depresss- contribut'' & AD & ``Vascular Demt- primary'' & -\\
``AD dem w/depresss- not contribut'' & AD & ``Vascular Demt- secondary'' & -\\
``AD dem w/oth (list B) contribut'' & AD & ``uncertain  possible NON AD dem'' & -\\
``AD dem w/oth (list B) not contrib'' & AD & ``uncertain dementia'' & -\\
``AD dem w/oth unusual feat/subs demt'' & AD & ``uncertain- possible NON AD dem'' & - \\
\end{longtable}
\end{small}

\subsubsection{Demographic Information}
A total of 590 subjects are involved in the current study, including 175 subjects with AD dementia and 415 normal controls. As shown in Table \ref{tab:demographic.info}, there is no significant difference on age ($t=1.5125$, $p > 0.05$), gender ($\chi^{2}=2.1782$, $p > 0.05$), and handness ($\chi^{2}=0.3926$, $p > 0.05$) between the two groups.

\begin{small}
\begin{table}[ht]
	\centering
	\caption{Demographic Information}
	\label{tab:demographic.info}
\begin{tabular}{l | c c c c}
\hline \hline
\multicolumn{1}{l}{} & AD Group & NC Group & Statistics & P-value \\
\hline \hline
\multicolumn{1}{l}{\makecell{Gender \\ (Male / Female)}} & $95 / 80$ & $196 / 219$ &  $2.1782^{a}$ & $0.1400$ \\
\hline
\multicolumn{1}{l}{\makecell{Age \\ (Mean $\pm$ SD)}} & $75.11\pm 7.67$ & $74.12 \pm 6.12$ &  $1.5125^{b}$ & $0.1316$ \\
\hline
\multicolumn{1}{l}{\makecell{Handness \\ (Left / Right / NA)}} & $18 / 154 / 3$ & $34 / 370 / 11$ & $0.3926^{a}$ & $0.5309$ \\
\hline \hline
\multicolumn{5}{l}{a: Welch's two-sample t-test} \\
\multicolumn{5}{l}{b: Pearson's chi-squared test with Yates' continuity correction} \\
\multicolumn{5}{l}{NA: Not applicable (indicating missing data)} \\
\end{tabular}
\end{table}
\end{small}

\begin{figure}[]
	\centering
    \includegraphics[width=0.9\linewidth]{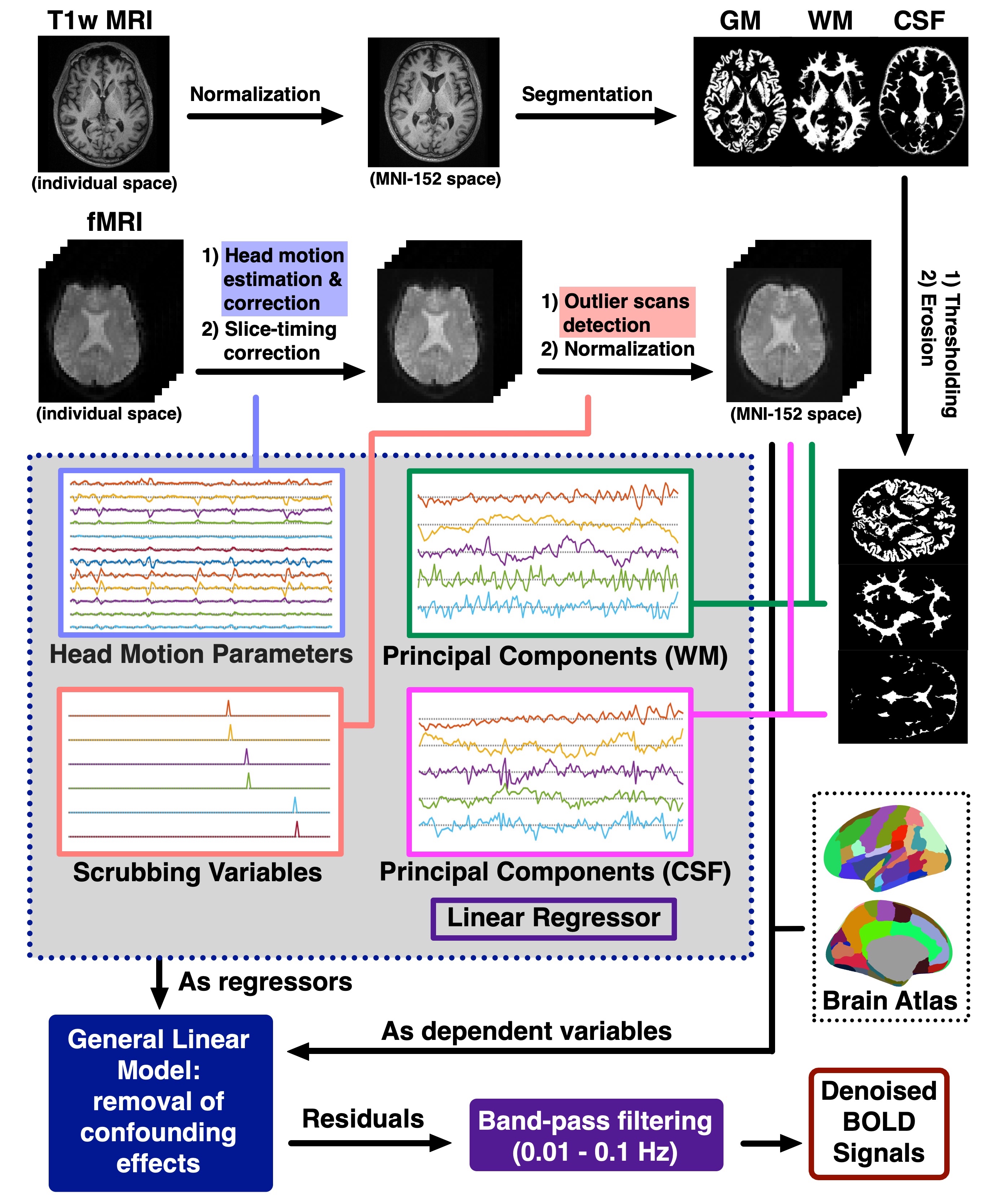}
	\caption[Processing Pipeline of Graph-theory-based Analysis] {{\bf Processing Pipeline of Graph-theory-based Analysis.} T1-weighted MR images are normalized into the MNI-152 space, and then segmented into grey matter (GM), white matter (WM), and cerebrospinal fluid (CSF). For BOLD fMRI, realignment (head motion estimation and correction), slice timing correction, outlier scans detection,  and normalization are applied. Various regressors are used  to remove confounding effects: (1) Principal components (PCs) from WM and CSF, (2) head motion parameters, (3) scrubbing variables, and (4) linear regressor. Finally, band-pass filtering ($0.01 - 0.1$) is applied.}
	\label{fig:pipeline.roi}
\end{figure}

\subsection{MR Images Processing} \label{section:pre-processing.roi}
MRI data were processed using the SPM12 and functional connectivity toolbox (CONN 18.b) \cite{whitfield2012conn}. The processing pipeline (Figure \ref{fig:pipeline.roi}) used in this study is mainly based on methods from SPM12. However, some other methods are applied in place of the ones from SPM12. For instance, a components-based method with anatomical noise ROI, aCompCor \cite{behzadi2007component}, is adopted to reduce the effects of noise from white matter (WM) and cerebrospinal fluid (CSF), instead of the global signal regression used in SPM12.

\subsubsection{Normalization and Segmentation of T1-weighted MRI}

\begin{figure}[]
	\centering
    \includegraphics[width=0.5\linewidth]{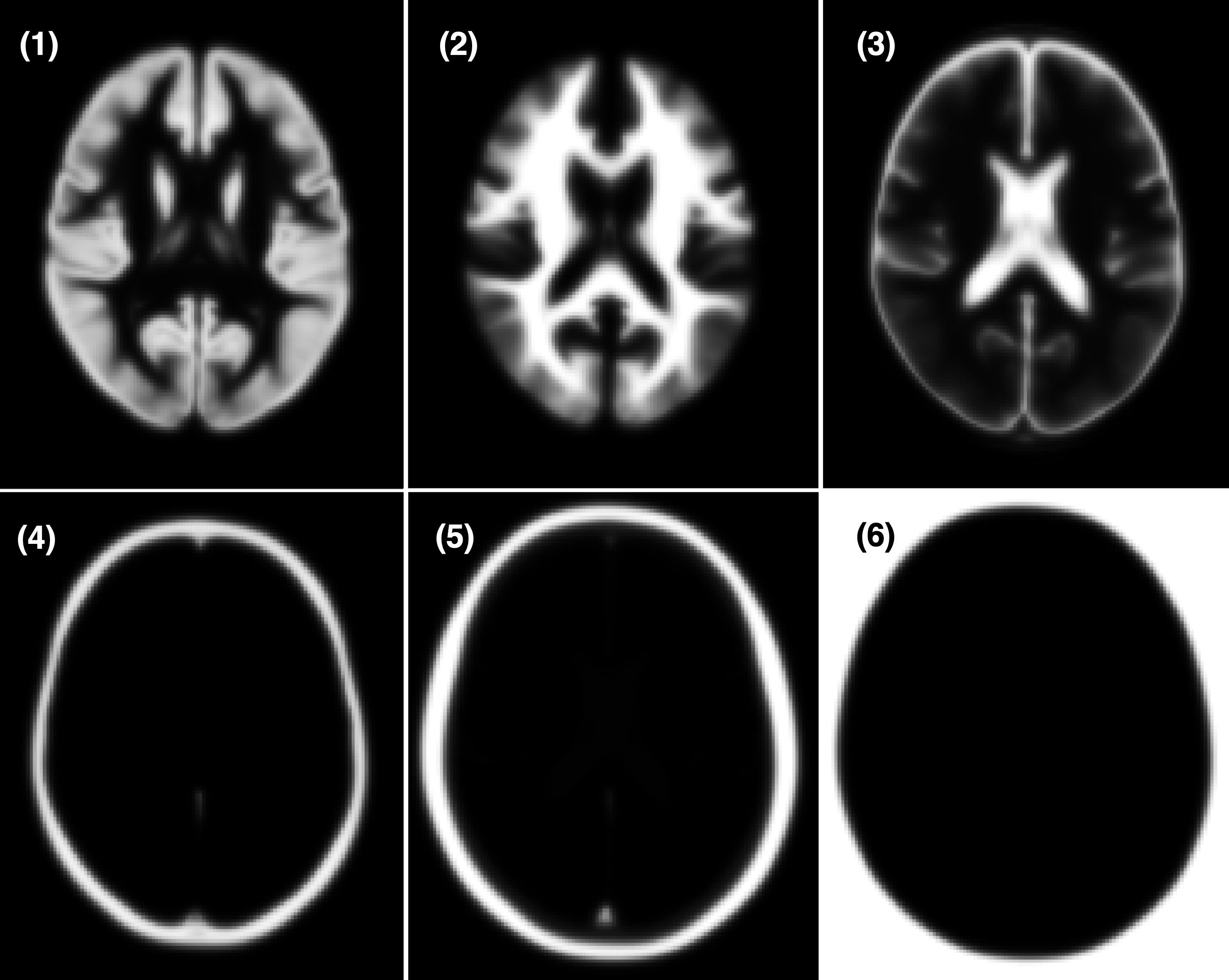}
	\caption[Tissue Probability Map] {{\bf Tissue Probability Map.} Defined in the MNI-152 space, the tissue probability map gives the prior probability for a voxel of being a specific brain tissue, including (1) grey matter, (2) white matter, (3) cerebrospinal fluid, (4) skull, (5) scalp, and (6) non-brain area. The values of these maps range $[0, 1]$. For one voxel, the probabilities sum up to 1.}
	\label{fig:TPM}
\end{figure}

Normalization of raw MR images into a common space is necessary for inter-subject comparison, and the MNI-152 space is the most widely used standard space. The MNI-152 space is defined by the prior tissue probability map (TPM) generated from structural MR images of 152 subjects. Specifically, first, all structural images are registered together; then, each voxel is assigned to one tissue type, resulting in individual tissue masks for each brain tissue; finally, the binary masks are averaged across all subjects to generate the tissue probability map. Therefore, various prior tissue probability maps (for each brain tissue), as a whole, represent the average shape and position of the brain in an MR image, which is the definition of a "standard space". The tissue probability map used is provided by SPM12 (Figure \ref{fig:TPM}), referred to as the TPM atlas.

For T1-weighted MR images, the first step applied is spatial normalization, which wraps the MRI volumes in individual spaces into the standard MNI-152 space. The spatial normalization consists of two steps: (1) estimating a non-linear deformation field that best overlays the TPM on the individual T1-weighted image; (2) wrapping the raw image with the inverse deformation estimated in step (1).

For segmentation, individual brain tissue probability maps for grey matter (GM), white matter (WM), and cerebrospinal fluid (CSF) are estimated. Then, a threshold is applied to each tissue probability map to generate a binary mask, which is further refined by morphology erosion. The eroded WM and CSF masks can serve as noise ROI masks, while the eroded GM mask is useful in restricting the subsequent analyses within the grey matter area.

\subsubsection{Head Motion Estimation and Correlation}
Head motion can introduce systematic co-variation across voxels, which increases the estimations of functional connections. More importantly, the distance-dependent signal modulation effect is reported \cite{power2012spurious} \cite{satterthwaite2012impact} \cite{van2012influence} in functional connectivity studies. It indicates that the variance added by motion artefact is similar in nearby voxels, resulting in a stronger short-distance correlation. Additionally, due to the application of head-motion-reduction processing methods, the observed long-distance correlation would often be decreased \cite{power2015recent}. Even a small head movement would cause signal disruption in BOLD fMRI \cite{power2015recent} and add spurious variance that can increase or decrease the observed functional connections. Thus, the estimation of head motion and the removal of head-motion-induced effects are crucial.

Among various artefacts, head motion is unique since it can be estimated from MR images via the realignment process (unlike, for example, the cardiac and respiratory artefacts that requires external recordings). Over the years, effort has been made to measure head motion, and considerable progress has been achieved. The simplest method is rigid body estimation, which measures the translation and rotation in the x, y, and z axes, producing 6 motion parameters. In 1996, an expansion of this motion estimation method was proposed by Karl Friston \cite{friston1996movement}. The expansion takes the form [$R R^{2} R_{t-1} R_{t-1}^{2}\dots R_{t-k} R_{t-k}^{2}$], where R denotes the 6 rigid body estimates, $R^{2}$ denotes their squares, and $R_{t-k}$ represents 6 parameters of the $k^{th}$ preceding time points. For instance, [$R R^{2} R_{t-1} R_{t-1}^{2}$] stands for a 24-parameter estimate, and [$R R^{2} R_{t-1} R_{t-1}^{2}R_{t-3} R_{t-3}^{2}$] refers to a 36-parameter estimate.

In this study, the 6-parameter rigid body realignment method is adopted, which is the strategy followed in other studies \cite{power2012spurious} \cite{satterthwaite2012impact} \cite{van2012influence}. These 6 parameters and their first-order derivatives, 12 parameters taking the form [$R R^\prime$] in total, are included in a linear regression model to regress out the head-motion-related variance in the following artefacts removal step. Although subsequent studies \cite{kundu2013integrated} have pointed out the inadequacy of these parameters, research by Jonathan Power and colleagues \cite{power2014methods} shows that more motion parameters cannot capture all head-movement-induced variance, and similar results were observed when 12, 24, and 36 motion parameters are included as regressors.

In the pre-processing pipeline of BOLD MR images, realignment is always the first step. Realignment is actually the head motion estimation and correction. It registers all other MR volumes in time series to a reference volume and exports the head motion parameters. The reference volume can be the volume at a specific time point, or the mean image averaged across time. In this study, the first volume is chosen as the reference. B-spline interpolation is used in the registration process. Head-motion-related variance can be partially reduced by realignment.

\subsubsection{Slice Timing Correction}
During the acquisition of a BOLD MR image with EPI sequence, a 3-D volume is actually a stack of 2-D slices collected one at a time. Therefore, for an fMRI volume at a time point, the voxel activations of each slice are not at the same time point. However, the ideal situation is that we can observe the activation of the whole brain at the same time. To resolve the conflict, slice timing correction can be used to interpolate slices to a reference slice. It has been shown to be an effective solution that can reliably increase sensitivity and effect power. The implementation details are as follows: (1) the acquisition time for each slice is read from the BIDS sidecar that comes with each NIfTI file; (2) all slices in a volume are interpolated to the slice acquired in the middle of time.

\subsubsection{Outlier Scans Detection} 
The Artifact Detection Tools (ART) are utilized for the automatic detection of volumes with severe head motion and global mean intensity change. The volumes that exceed the specified threshold are considered outlier scans and labelled by scrubbing variables. The ART-based outlier detection function is conveniently integrated into the CONN toolbox. The "intermediate settings," which treat the $97^{th}$ percentiles in the normative sample as the threshold, are selected. At the end of this step, a set of scrubbing variables is stored for further use as regressors in the general linear model.

\subsubsection{Normalization of BOLD MRI}
In the CONN toolbox, spatial normalization of BOLD MR images can be implemented using two methods: direct and indirect normalization.
\paragraph{Direct normalization} It involves normalizing the T1-weighted MR images and BOLD MR images separately. It is implemented in the same way as spatial normalization of T1w MR images: (1) first, a nonlinear deformation field that wraps the TPM atlas to the individual BOLD MR image is estimated; (2) then, the inverse transformation is applied to normalize the BOLD MR image into MNI-152 space.

\paragraph{Indirect normalization} It involves two steps: (1) first, the BOLD MR image is co-registered to the T1-weighted MR image, generating a transformation matrix $T_{1}$; (2) then, the obtained transformation matrix $T_{1}$ is multiplied by the transformation matrix $T_{2}$, which wraps the T1-weighted image to MNI-152 space. Finally, the BOLD image can be transformed into the standard space with the matrix $T_{2}T_{1}$.

As a step in the default pre-processing pipeline of the CONN toolbox, the direct normalization scheme is applied.

\subsubsection{Spatial Smoothing of BOLD MRI}
Spatial smoothing with a Gaussian kernel is typically the final step in a spatial processing pipeline for a BOLD MR image. However, it is not utilized in this study because the current study only involves ROI-based graph analyses. That is, the final processed and de-noised signal is the mean time series of an ROI. Spatial smoothing prior to averaging would not significantly alter the resulting mean signal. We only consider spatial smoothing to be a necessary image processing step when it is followed by voxel-based analyses. This approach is also used by the CONN toolbox.

\subsubsection{Artefacts Removal} \label{section:graph.denoising}
The removal of artefacts is a critical step in BOLD signal processing, aimed at mitigating or eliminating the confounding effects of non-neuronal oscillations caused by head movement, cardiac pulsation, respiratory motion, and other systematic noises. Without this step, it is challenging for researchers to determine whether the findings are genuine or simply driven by artefacts. A general linear model (GLM) is used for artefacts removal, with mean BOLD signals extracted from ROIs defined by a prior brain atlas, and a variety of variables defined as regressors of the GLM (as shown in Figure \ref{fig:pipeline.roi}).

In the previous realignment step, the nuisance head motion effect can be partially mitigated by interpolation. Further, this confounding effect is reduced by regressing out 12 head motion parameters (including 3 translation parameters, 3 rotation parameters, and their first-order derivatives) and scrubbing variables. One effect of head motion is increased short-distance correlations and decreased long-distance correlations. Interestingly, it has been reported that scrubbing high-motion frames from fMRI data can decrease short-term correlations and increase long-term correlations \cite{power2012spurious}, indicating that the variance originating from head motion can be explained by scrubbing variables.

To remove other nuisance effects, the aCompCor method \cite{behzadi2007component} is utilized, which is a component-based method with anatomical noise ROIs. Specifically, WM and CSF masks are used to define the WM and CSF areas as noise ROIs. Then, five principal components (PCs) for each noise ROI are calculated via principal component analysis. Lastly, the five PCs from WM and five PCs from CSF are entered into the linear model as regressors. Additionally, the linear trend is removed by adding a linear regressor into the GLM. Finally, the residual time series are band-pass filtered at [$0.01 - 0.1$] Hz to retain neuroactivity-related intrinsic signal fluctuations.

\subsection{Functional Brain Network Construction}
\subsubsection{Nodes Definition}
\begin{figure}[]
	\centering
	\includegraphics[width=0.9\linewidth] {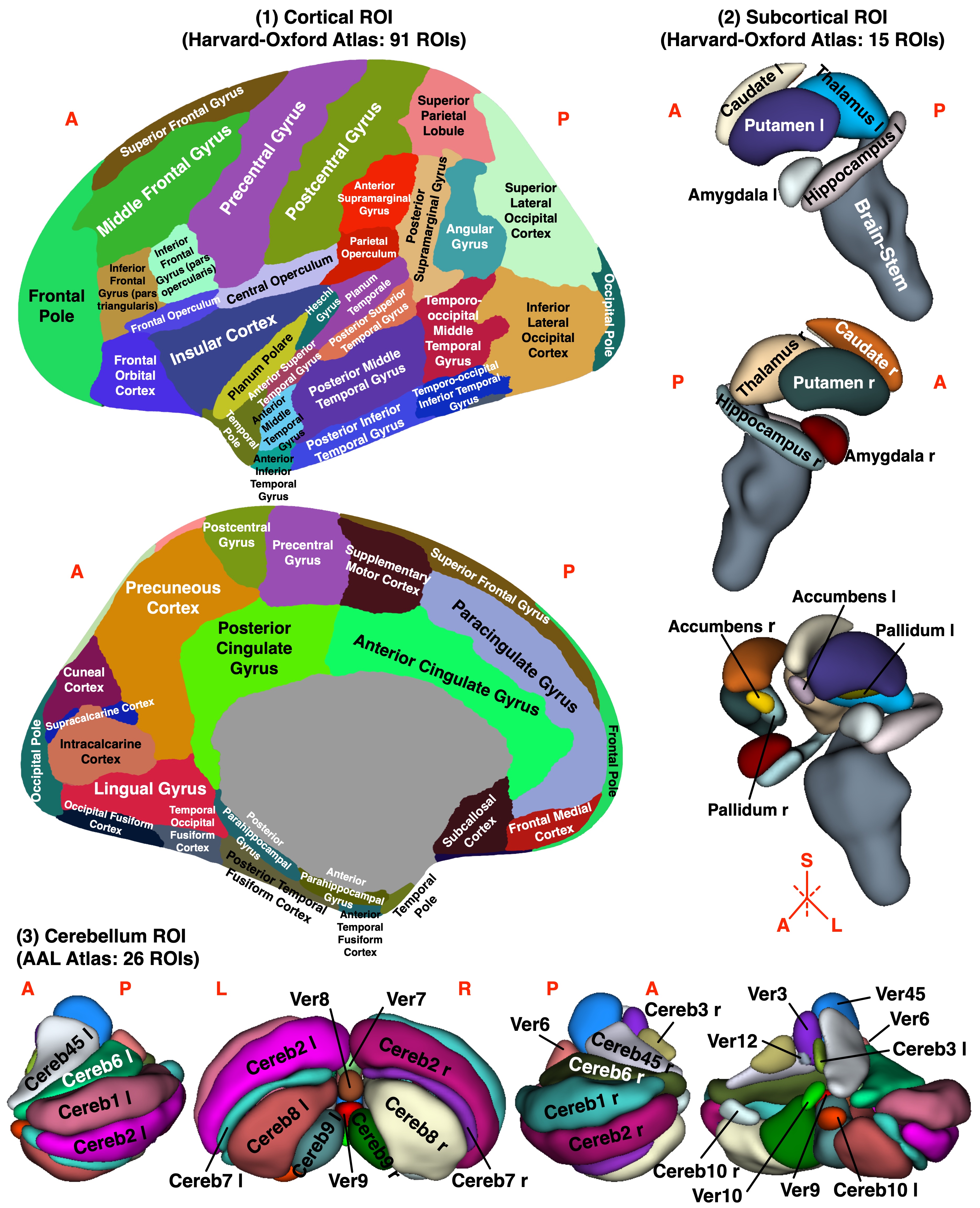}
	\caption[Brain Atlas] {{\bf Brain Atlas.} It is a customized brain regions parcellation scheme, combining (1) 91 cortical regions (from the Harvard-Oxford Atlas), (2) 15 subcortical structures (from the Harvard-Oxford Atlas), and (3) 26 cerebellum regions (from the AAL Atlas). In the 3-D rendering of subcortical and cerebellum regions, spatial smoothing is applied for better visualization. (A: anterior; P: posterior; L: left; R: right; S: superior)}
	\label{fig:brain.atlas}
\end{figure}

\begin{figure}[]
	\centering
	\includegraphics[width=0.9\linewidth] {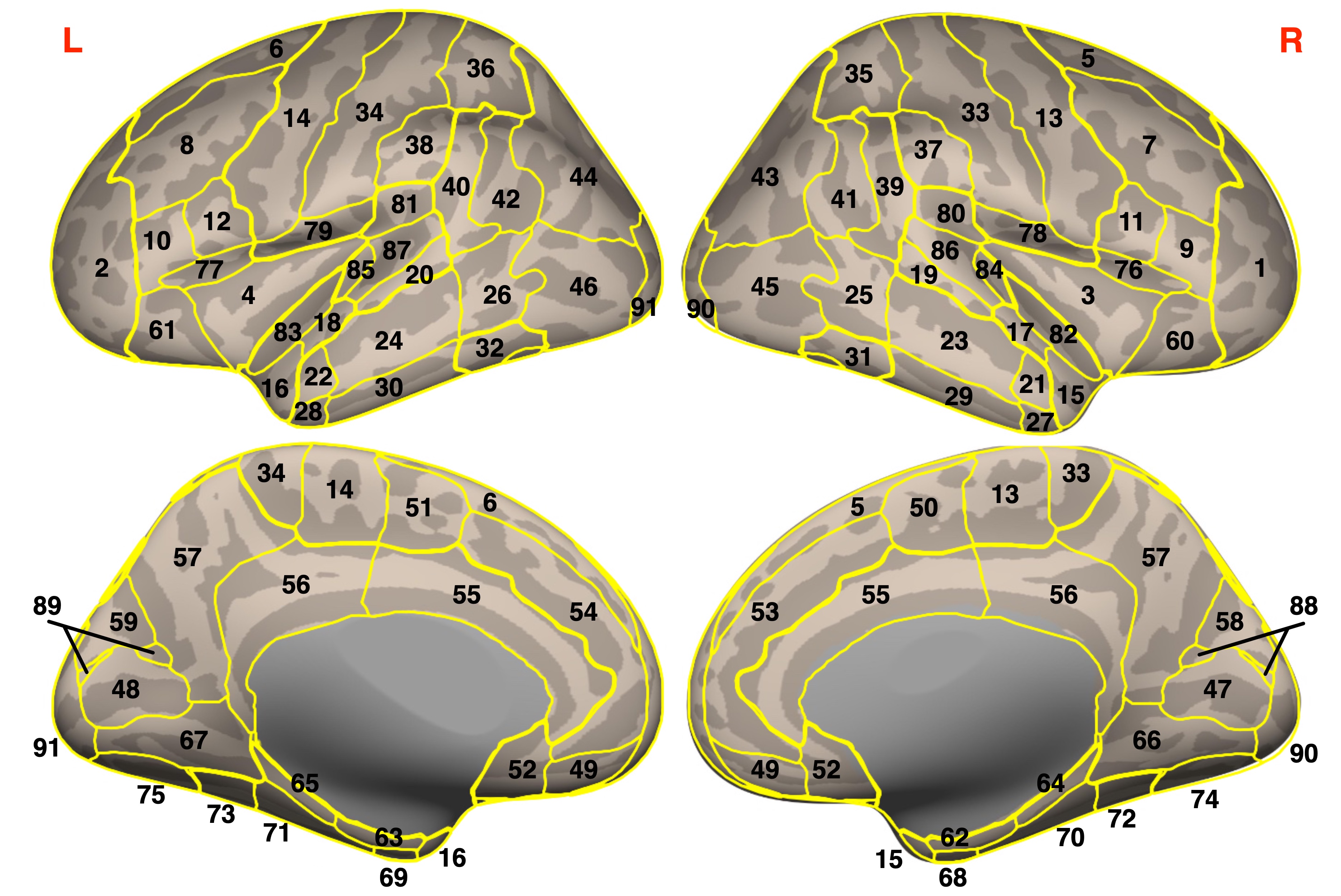}
	\caption[Contours of Cortical Regions] {{\bf Contours of Cortical Regions.} The contours of 91 cortical regions defined in the Harvard-Oxford Atlas. Number is the ID for each region (Table \ref{tab:atlas.info}).}
	\label{fig:brain.atlas.contour}
\end{figure}

In functional brain network analyses, nodes are typically defined as brain regions that are structurally or functionally distinct. However, delineating the boundaries of brain regions is a complex and continually evolving research field. Over the past few decades, numerous atlases have been developed, including the Automated Anatomical Labeling (AAL) atlas \cite{rolls2015implementation,tzourio2002automated}, the 7- and 17-Network Parcellation \cite{thomas2011organization}, the 400-, 600-, 800-, and 1000-area parcellation \cite{schaefer2017local}, the Atlas of Intrinsic Connectivity of Homotopic Areas (AICHA) \cite{joliot2015aicha}, the Human Connectome Project Multi-Modal Parcellation (HCP-MMP) \cite{glasser2016multi}, the Cortical Area Parcellation \cite{gordon2014generation} released by a group in Washington University, and the Human Brainnetome Atlas \cite{fan2016human}.
In the current study, we use a customized atlas named the ``Harvard-Oxford-AAL'' atlas, which includes 91 cortical ROIs and 15 subcortical ROIs from the Harvard-Oxford Atlas distributed with the FMRIB Software Library (FSL), as well as 26 cerebellum ROIs from the AAL atlas. An illustration of all 132 ROIs can be found in Figure \ref{fig:brain.atlas}, and Table \ref{tab:atlas.info} provides additional details. Figure \ref{fig:brain.atlas.contour} displays the contours of cortical regions.

\begin{center}
\begin{small}
\setlength\LTleft{0 cm}
\setlength\LTright{0 cm}
\begin{longtable}[c] {l | ll}
	\caption{ID, Names, and Abbreviations of Brain Regions} \label{tab:atlas.info} \\
\hline  \hline 
\multicolumn{1}{l}{\textbf{ID}} & 
\multicolumn{1}{l}{\textbf{Abbrev.}} & 
\multicolumn{1}{l}{\textbf{Brain Region}} \\ 
\hline \hline
\endfirsthead
\multicolumn{3}{l}
{{\tablename\ \thetable{} -- continued from previous page}} \\
\hline 
\multicolumn{1}{l}{\textbf{ID}} & 
\multicolumn{1}{l}{\textbf{Abbrev.}} & 
\multicolumn{1}{l}{\textbf{Brain Region}}  \\ 
\hline 
\endhead
\hline \multicolumn{3}{r}{{Continued on next page}} \\ \hline
\endfoot
\hline \hline
\endlastfoot
1     & FP r  & Frontal Pole Right \\
2     & FP l  & Frontal Pole Left \\
3     & IC r  & Insular Cortex Right \\
4     & IC l  & Insular Cortex Left \\
5     & SFG r  & Superior Frontal Gyrus Right \\
6     & SFG l  & Superior Frontal Gyrus Left \\
7     & MidFG r  & Middle Frontal Gyrus Right \\
8     & MidFG l  & Middle Frontal Gyrus Left \\
9     & IFG tri r  & Inferior Frontal Gyrus, pars triangularis Right \\
10    & IFG tri l  & Inferior Frontal Gyrus, pars triangularis Left \\
11    & IFG oper r  & Inferior Frontal Gyrus, pars opercularis Right \\
12    & IFG oper l  & Inferior Frontal Gyrus, pars opercularis Left \\
13    & PreCG r  & Precentral Gyrus Right \\
14    & PreCG l  & Precentral Gyrus Left \\
15    & TP r  & Temporal Pole Right \\
16    & TP l  & Temporal Pole Left \\
17    & aSTG r  & Superior Temporal Gyrus, anterior division Right \\
18    & aSTG l  & Superior Temporal Gyrus, anterior division Left \\
19    & pSTG r  & Superior Temporal Gyrus, posterior division Right \\
20    & pSTG l  & Superior Temporal Gyrus, posterior division Left \\
21    & aMTG r  & Middle Temporal Gyrus, anterior division Right \\
22    & aMTG l  & Middle Temporal Gyrus, anterior division Left \\
23    & pMTG r  & Middle Temporal Gyrus, posterior division Right \\
24    & pMTG l  & Middle Temporal Gyrus, posterior division Left \\
25    & toMTG r  & Middle Temporal Gyrus, temporooccipital part Right \\
26    & toMTG l  & Middle Temporal Gyrus, temporooccipital part Left \\
27    & aITG r  & Inferior Temporal Gyrus, anterior division Right \\
28    & aITG l  & Inferior Temporal Gyrus, anterior division Left \\
29    & pITG r  & Inferior Temporal Gyrus, posterior division Right \\
30    & pITG l  & Inferior Temporal Gyrus, posterior division Left \\
31    & toITG r  & Inferior Temporal Gyrus, temporooccipital part Right \\
32    & toITG l  & Inferior Temporal Gyrus, temporooccipital part Left \\
33    & PostCG r  & Postcentral Gyrus Right \\
34    & PostCG l  & Postcentral Gyrus Left \\
35    & SPL r  & Superior Parietal Lobule Right \\
36    & SPL l  & Superior Parietal Lobule Left \\
37    & aSMG r  & Supramarginal Gyrus, anterior division Right \\
38    & aSMG l  & Supramarginal Gyrus, anterior division Left \\
39    & pSMG r  & Supramarginal Gyrus, posterior division Right \\
40    & pSMG l  & Supramarginal Gyrus, posterior division Left \\
41    & AG r  & Angular Gyrus Right \\
42    & AG l  & Angular Gyrus Left \\
43    & sLOC r  & Lateral Occipital Cortex, superior division Right \\
44    & sLOC l  & Lateral Occipital Cortex, superior division Left \\
45    & iLOC r  & Lateral Occipital Cortex, inferior division Right \\
46    & iLOC l  & Lateral Occipital Cortex, inferior division Left \\
47    & ICC r  & Intracalcarine Cortex Right \\
48    & ICC l  & Intracalcarine Cortex Left \\
49    & MedFC  & Frontal Medial Cortex \\
50    & SMA r  & Supplementary Motor Cortex Right \\
51    & SMA L & Supplementary Motor Cortex Left \\
52    & SubCalC  & Subcallosal Cortex \\
53    & PaCiG r  & Paracingulate Gyrus Right \\
54    & PaCiG l  & Paracingulate Gyrus Left \\
55    & AC    & Cingulate Gyrus, anterior division \\
56    & PC    & Cingulate Gyrus, posterior division \\
57    & Precuneous  & Precuneous Cortex \\
58    & Cuneal r  & Cuneal Cortex Right \\
59    & Cuneal l  & Cuneal Cortex Left \\
60    & FOrb r  & Frontal Orbital Cortex Right \\
61    & FOrb l  & Frontal Orbital Cortex Left \\
62    & aPaHC r  & Parahippocampal Gyrus, anterior division Right \\
63    & aPaHC l  & Parahippocampal Gyrus, anterior division Left \\
64    & pPaHC r  & Parahippocampal Gyrus, posterior division Right \\
65    & pPaHC l  & Parahippocampal Gyrus, posterior division Left \\
66    & LG r  & Lingual Gyrus Right \\
67    & LG l  & Lingual Gyrus Left \\
68    & aTFusC r  & Temporal Fusiform Cortex, anterior division Right \\
69    & aTFusC l  & Temporal Fusiform Cortex, anterior division Left \\
70    & pTFusC r  & Temporal Fusiform Cortex, posterior division Right \\
71    & pTFusC l  & Temporal Fusiform Cortex, posterior division Left \\
72    & TOFusC r  & Temporal Occipital Fusiform Cortex Right \\
73    & TOFusC l  & Temporal Occipital Fusiform Cortex Left \\
74    & OFusG r  & Occipital Fusiform Gyrus Right \\
75    & OFusG l  & Occipital Fusiform Gyrus Left \\
76    & FO r  & Frontal Operculum Cortex Right \\
77    & FO l  & Frontal Operculum Cortex Left \\
78    & CO r  & Central Opercular Cortex Right \\
79    & CO l  & Central Opercular Cortex Left \\
80    & PO r  & Parietal Operculum Cortex Right \\
81    & PO l  & Parietal Operculum Cortex Left \\
82    & PP r  & Planum Polare Right \\
83    & PP l  & Planum Polare Left \\
84    & HG r  & Heschl's Gyrus Right \\
85    & HG l  & Heschl's Gyrus Left \\
86    & PT r  & Planum Temporale Right \\
87    & PT l  & Planum Temporale Left \\
88    & SCC r  & Supracalcarine Cortex Right \\
89    & SCC l  & Supracalcarine Cortex Left \\
90    & OP r  & Occipital Pole Right \\
91    & OP l  & Occipital Pole Left \\
92    & Thalamus r & Thalamus Right \\
93    & Thalamus l & Thalamus Left \\
94    & Caudate r & Caudate Right \\
95    & Caudate l & Caudate Left \\
96    & Putamen r & Putamen Right \\
97    & Putamen l & Putamen Left \\
98    & Pallidum r & Pallidum Right \\
99    & Pallidum l & Pallidum Left \\
100   & Hippocampus r & Hippocampus Right \\
101   & Hippocampus l & Hippocampus Left \\
102   & Amygdala r & Amygdala Right \\
103   & Amygdala l & Amygdala Left \\
104   & Accumbens r & Accumbens Right \\
105   & Accumbens l & Accumbens Left \\
106   & Brain-Stem & Brain-Stem \\
107   & Cereb1 l  & Cerebellum Crus1 Left \\
108   & Cereb1 r  & Cerebellum Crus1 Right \\
109   & Cereb2 l  & Cerebellum Crus2 Left \\
110   & Cereb2 r  & Cerebellum Crus2 Right \\
111   & Cereb3 l  & Cerebellum 3 Left \\
112   & Cereb3 r  & Cerebellum 3 Right \\
113   & Cereb45 l  & Cerebellum 4 5 Left \\
114   & Cereb45 r  & Cerebellum 4 5 Right \\
115   & Cereb6 l  & Cerebellum 6 Left \\
116   & Cereb6 r  & Cerebellum 6 Right \\
117   & Cereb7 l  & Cerebellum 7b Left \\
118   & Cereb7 r  & Cerebellum 7b Right \\
119   & Cereb8 l  & Cerebellum 8 Left \\
120   & Cereb8 r  & Cerebellum 8 Right \\
121   & Cereb9 l  & Cerebellum 9 Left \\
122   & Cereb9 r  & Cerebellum 9 Right \\
123   & Cereb10 l  & Cerebellum 10 Left \\
124   & Cereb10 r  & Cerebellum 10 Right \\
125   & Ver12  & Vermis 1 \& Vermis 2 \\
126   & Ver3  & Vermis 3 \\
127   & Ver45  & Vermis 4 \& Vermis 5 \\
128   & Ver6  & Vermis 6 \\
129   & Ver7  & Vermis 7 \\
130   & Ver8  & Vermis 8 \\
131   & Ver9  & Vermis 9 \\
132   & Ver10  & Vermis 10\\
\end{longtable}
\end{small}
\end{center}

\subsubsection{Edges Definition}
In weightless bidirectional graph analysis, defining the edges involves two steps: (1) determining the connectivity measure, and (2) applying a threshold to create weightless connections.

In functional brain network analysis using BOLD fMRI data, there are two major types of connectivity measures: correlation measures and regression measures. Correlation measures utilize a correlation coefficient to quantify the statistical relationship between two BOLD signals, such as Pearson’s linear correlation coefficient, intra-class correlation, or rank-based correlation coefficients like Spearman’s rank correlation coefficient and Kendall tau rank correlation coefficient. Various correlation measures have their own characteristics and usability, but they all assume values ranging from $-1$ to $+1$, where $\pm1$ indicates the strongest relationship, and 0 represents the weakest relationship. Regression measures, on the other hand, use a regression coefficient to measure the relationship between variables. Regression measures offer more flexibility, such as when analyzing the relationship between two variables using bivariate regression measures or controlling for all other factors using multivariate regression measures.

In this study, the functional connectivity between two ROIs is defined using Pearson’s linear correlation coefficient, which is further subjected to Fisher z-transformation, a standard step in functional connectivity analyses for hypothesis testing of the population correlation coefficient. To construct a weightless graph from these weighted connections, thresholding is necessary. Various thresholding methods exist, categorized into absolute and relative thresholding methods. Absolute thresholding applies a fixed threshold to connectivity measures, while relative thresholding retains a specific proportion of connections within a graph. In this study, relative thresholding is used, retaining only the top $15\%$ connections.

In fMRI connectivity analysis, negative connections (anti-correlations) can be observed using linear correlation coefficients. Whether these negative connections are spurious or physiologically meaningful is an open question. To investigate their effects on the results, three parallel experiments were conducted: exploring the properties of the negative functional network by retaining only anti-correlations, exploring the properties of the positive functional network by eliminating all anti-correlations and exploring the properties of the mixed functional network using absolute correlation values.

\subsection{Graph-Theory-Based Network Properties}
\subsubsection{Degree and Cost}
\paragraph{Nodal Degree and Cost}
Two ways to quantify the centrality of each node in a graph are measuring the number and proportion of edges directly linked to that node – that is, calculating the degree and cost respectively. Degree and cost are similarly defined and related, both of which are meant to characterize the local connectivity of each node. Here, local connectivity depicts how densely one node is connected to all other nodes in the same graph. If one node is densely connected to others, then its degree or cost will be high. On the contrary, if one node is sparsely connected, its degree or cost will be low. 

\begin{itemize}
    \item \emph{Nodal degree} represents the number of edges directly connected to a node and is defined as
        \begin{equation}
        K_{i} = \sum_{i\ne j \in N} a_{ij},
        \end{equation}
        where $K_{i}$ is the degree of node $i$, $N$ stands for the set of all nodes in a graph, and $a$ is a binary value (0 for non-connection, 1 for connection) indicating the connectivity, in other words, the connection status.
    \item \emph{Nodal cost} represents the proportion of direct edges among all possible connections and is defined as 
        \begin{equation}
        C_{i} = \frac{1}{n-1} \sum_{i\ne j \in N}a_{ij},
        \end{equation}
        where $C_{i}$ is the cost of node $i$, $N$ is the set of all nodes, $a_{ij}$ represents connectivity, and $n$ is the number of nodes in the graph. 
\end{itemize}

Actually, it is obvious that the equation $C_{i} = \frac{K_{i}}{n-1}$ holds. The major reason why cost is defined in addition to degree is that the variation derived from the size of the graph needs to be corrected when the cross-graph comparison is performed. For example, directly comparing the degree of a node in a 90-nodes-graph and that in a 264-nodes-graph is meaningless since the former is most likely smaller than the latter. Yet, the cost of a node can be used for direct comparison in this scenario. When degree and cost refer to the individual properties of a node, we usually call it nodal degree and nodal cost, or the degree of a node and the cost of a node.

\paragraph{Mean Degree and Cost}
From the nodal property, we can calculate the global property of the whole graph -- the degree or cost of a graph, which is simply the average of nodal degree or nodal cost. The degree and cost of a graph are also called mean degree and mean cost. While nodal degree and cost are metrics for measuring nodal centrality, mean degree and cost are measures of network integration, with larger values denoting a higher level of integration, and vice versa.

\begin{itemize}
    \item \emph{Mean degree} is defined as
        \begin{equation}
        K = \frac{1}{n} \sum_{i \in N}K_{i} = \frac{1}{n} \sum_{i\in N} \sum_{i\ne j \in N } a_{ij}.
        \end{equation}
    \item \emph{Mean cost} is defined as
        \begin{equation}
        C = \frac{1}{n} \sum_{i\in N}C_{i} = \frac{1}{n} \sum_{i\in N} \frac{\sum_{i\ne j\in N} a_{ij}}{n-1}.
        \end{equation}
\end {itemize}

\subsubsection{Shortest Path Length}
\paragraph{Average Shortest Path Length}
In a binary bidirectional graph, all edges (\emph{i.e.}, connections or links) are weightless and bidirectional. Two nodes within a network can be directly or indirectly connected, with many possible paths between them. The shortest path length between two nodes is defined as the minimum distance of all possible paths connecting them. If no path exists between the two nodes, the shortest path length is defined as $+\infty$.
The average shortest path length of all paths emitted from a node quantitatively describes the centrality of that node and can be used as a measure of integration. It is calculated as
\begin{equation}
L_i = \frac{1}{n-1}\sum_{i\neq j\in N} d_{ij},
\end{equation}
where $L_i$ is the average shortest path length of node $i$, $d_{ij}$ is the shortest path length between node $i$ and $j$, $N$ is the set of all nodes, and $n$ is the number of nodes in the graph. Like nodal degree and cost, the average shortest path length is a measure of nodal centrality, with a smaller value indicating a higher level of centrality. However, the definition of infinite distance means that if a node is disconnected from node $i$, the average shortest path length of node $i$ becomes infinite. This is referred to as "the infinity property of the average shortest path length" in this article. It should be noted that this is a drawback of the definition of the average shortest path length because any disconnected node will render this metric infinite.

\paragraph{Characteristic Path Length}
The average shortest path length is a measure of nodal property in a graph, and its corresponding global property is called the characteristic path length of a graph \cite{watts1998collective}. It is defined as 
\begin{equation}
L=\frac{1}{n} \sum_{i \in N},
\end{equation}
\begin{equation}
L_{i}=\frac{1}{n} \sum _{i \in N}\frac{ \sum_{i \neq j \in N}d_{ij}}{n-1},
\end{equation}
which is simply the mean of the average shortest path length across all nodes in a graph. Like mean degree and cost, the characteristic path length is a measure of network integration. The smaller its value, the more integrated the network is. An integrated network implies that, on average, each node has short paths to all other nodes, meaning that the nodes are closely linked to one another.
However, due to the infinity property of the average shortest path length, any isolated node in the graph will also make the characteristic path length infinite. Therefore, its application is restricted to connected graphs. To overcome this limitation, researchers have defined metrics not based on the shortest path length, such as global efficiency.

\subsubsection{Clustering Coefficient}
\paragraph{Nodal Clustering Coefficient}
In a graph, the neighbours of node $i$ are defined as all nodes that directly connect to node $i$. In the analysis of small-world networks, the clustering coefficient of node $i$ is defined by Watts and colleagues \cite{watts1998collective} as the fraction of edges among all the possible edges in the subgraph of its neighbours. It is calculated as 
\begin{equation}
CC_{i}=\frac{ \sum_{p,q \in N{i}}a_{pq}}{K_{i} \left( K_{i}-1 \right) },
\end{equation}
where $CC_{i}$ denotes the clustering coefficient of node $i$, $N_{i}$ denotes the subgraph of its neighbours, $a_{pq}$ indicates connectedness, and $K_{i}$ is the degree of node $i$. As a measure of nodal property, the clustering coefficient of a node is also referred to as the nodal clustering coefficient. It characterizes the cliquishness or clustering property of a node.

\paragraph{Mean Clustering Coefficient}
Similarly, the clustering coefficient of a graph is defined as the average of all nodal clustering coefficients: 
\begin{equation}
CC=\frac{1}{n} \sum _{i \in N}CC_{i}=\frac{1}{n} \sum _{i \in N}\frac{ \sum _{p,q \in N_{i}}a_{pq}}{K_{i} \left( K_{i}-1 \right) }.
\end{equation}
It is sometimes referred to as the mean clustering coefficient. Unlike mean degree, mean cost, and characteristic path length, the mean clustering coefficient is used to measure the segregation of a network. It reflects, on average, the extent of local connectivity around a node.

\subsubsection{Efficiency}
The existence of the infinity property means that the rational application of the average shortest path length or characteristic path length is subject to the connectivity of the graph, which requires that all nodes in the graph are connected. To address this limitation, Latora and colleagues \cite{latora2001efficient} proposed the concept of efficiency to describe the behaviour of a network. As its name suggests, efficiency is meant to quantify how efficient the communication or information exchange between nodes is. Different papers use various terminologies to refer to the concept of efficiency in graph-theory-based metrics, which can lead to confusion. In this article, we provide a brief explanation of the terms used. First, there are two classes of metrics - one for nodal property and the other for the overall property of the whole graph. Some may use "local" and "global" to indicate nodal and graph-level properties. However, "local" and "global" have specific meanings in the definitions of efficiency - local efficiency and global efficiency are two different metrics. Therefore, when discussing efficiency, the terms "local" and "global" solely indicate the types of efficiency, but not the scope. Second, to clarify the scope of a metric, we state it clearly in the name. For example, "global efficiency of a node" indicates that it is a nodal property, and the measure here is "global efficiency". Another example is that "local efficiency of a graph" refers to graph-level local efficiency. We avoid using terms like "nodal local efficiency" or "nodal global efficiency" to prevent confusion.

\paragraph{Global Efficiency of a Node}
Efficiency is closely related to the shortest path length by definition. Let us assume that the transfer of information in a network is parallel, i.e., the signal from a node is sent concurrently through all edges. As a result, the efficiency of communication between node $i$ and $j$ is determined by the shortest path between them. The efficiency is then defined as the inverse of the shortest path length: $\epsilon_{ij}=\frac{1}{d_{ij}}$, where $\epsilon_{ij}$ is the efficiency of information transfer and $d_{ij}$ is the shortest path length. The global efficiency of node $i$ is defined as
\begin{equation}
E_{i, glob}=\frac{1}{n-1} \sum_{i \neq j \in N}\epsilon_ {ij}.
\end{equation}
With this definition, the global efficiency of a node can be meaningfully calculated on a disconnected graph, where isolated nodes exist. For example, when node $i$ and $j$ are disconnected, $d_{ij}= +\infty$ and $\epsilon_{ij}=0$, and therefore, $E_{i, glob}$ or $E_{j, glob}$ is not affected by the lack of connection.

\paragraph{Global Efficiency of a Graph}
We can calculate the global efficiency of a graph, which is a measure of the whole network, by averaging the global efficiency of all nodes. The global efficiency of a node is the reciprocal of the average shortest path length between the node and all other nodes in the network. If a graph is disconnected, the global efficiency is still valid as it is not affected by disconnection. Specifically, the global efficiency of a graph is defined as follows:
\begin{equation}
E_{glob}=\frac{1}{n} \sum_{i \in N}E_{i, glob}=\frac{1}{n} \sum _{i \in N}\frac{ \sum _{i \neq j \in N}\epsilon _{ij}}{n-1}=\frac{1}{n} \sum _{i \in N}\frac{ \sum_{i \neq j \in N}d_{ij}^{-1}}{n-1}.
\end{equation}

\begin{figure}[]
	\centering
    \includegraphics[width=0.8\linewidth] {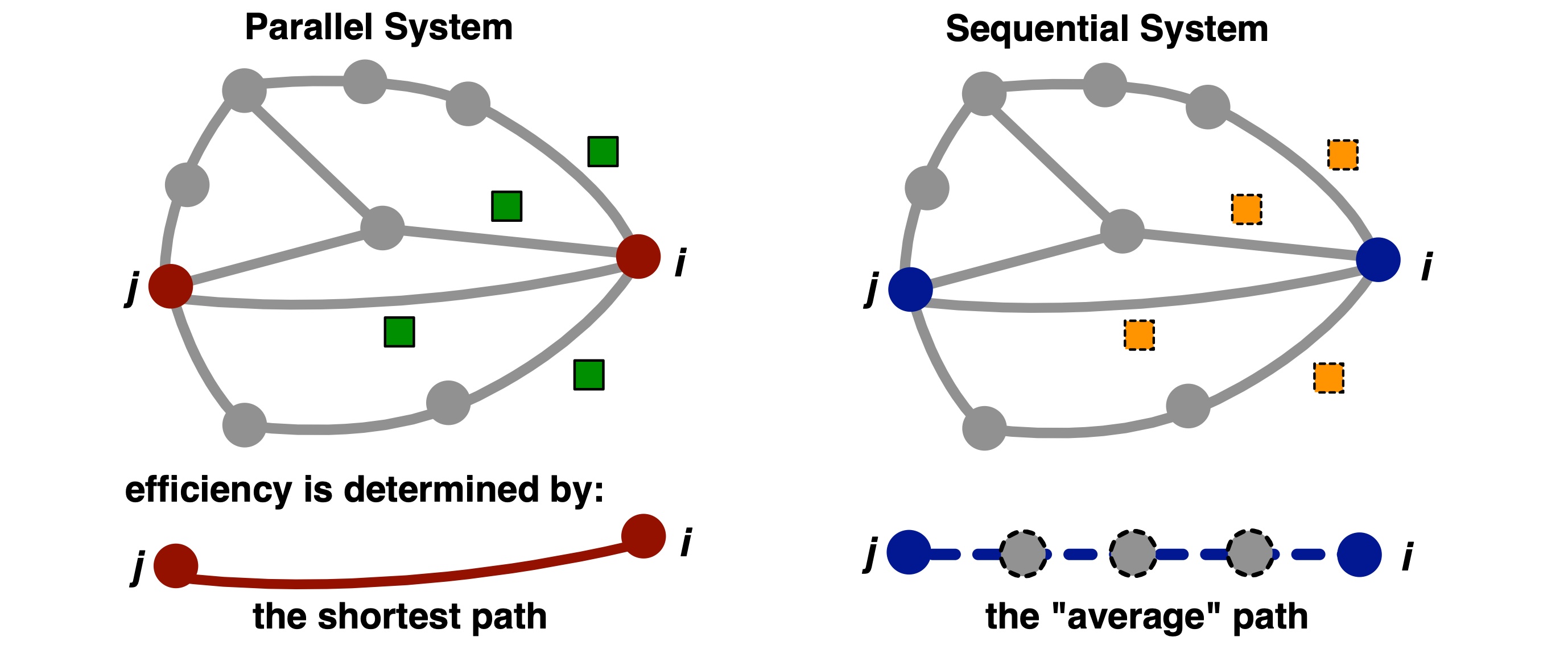}
	\caption[Parallel and Sequential System] {{\bf Parallel and Sequential System.} In a parallel system (left column), the delivery of information from node $i$ to node $j$ occurs concurrently through every path, making the shortest path length the bottleneck of system efficiency. In a sequential system (right column),  the transmission of information can be considered as sending a packet through one random path at a time. Therefore, the system efficiency is determined by the imaginary ``average path’’. A parallel system is not affected by long paths while a sequential system is.}
	\label{fig:parallel.sequential.sys}
\end{figure}

The global efficiency and characteristic path length are both based on the shortest path length between nodes and serve as evaluation indices of network integration. A highly integrated network has a high global efficiency or a short characteristic path length.
The difference between global efficiency and characteristic path length can be understood through an analogy provided by the original paper \cite{latora2001efficient}. Global efficiency is like a parallel system where information is sent simultaneously through all paths, while the characteristic path length is like a sequential system where information is transferred through one path at a time.
To further illustrate the analogy, we can define efficiency as the number of packets transferred in a time unit. In a parallel system (left column in Figure \ref{fig:parallel.sequential.sys}), packets are sent concurrently through all paths, and the shortest path length determines the efficiency. In a sequential system (right column in Figure \ref{fig:parallel.sequential.sys}), packets are transferred through one path at a time, and the efficiency is the number of packets successfully transferred through an imaginary "average path" in a time unit. Long paths decrease the efficiency by lengthening the "average path".

\paragraph{Local Efficiency of a Node}
The global efficiency of a node measures its communication with all other nodes in the graph, while the local efficiency quantifies the global efficiency of a subgraph consisting only of its neighbours. Specifically, the local efficiency of node $i$ is defined as the global efficiency of its subgraph of neighbours: 
\begin{equation}
E_{i, local}=\frac{1}{K_{i}} \sum _{p \in N_{i}}\frac{ \sum _{p \neq q \in N_{i}} \epsilon _{pq}}{K_{i}-1}=\frac{ \sum _{p \neq q \in N_{i}} d_{pq}^{-1}}{K_{i} \left( K_{i}-1 \right) },
\end{equation}
where $K_{i}$ is the degree of node $i$, $N_{i}$ is the subgraph of its neighbours, $\epsilon_{pq}$ represents the efficiency of the shortest path between node $p$ and $q$, and $d_{pq}$ is the shortest path length.
		
\paragraph{Local Efficiency of a Graph}
The local efficiency of a graph is the average local efficiency of all subgraphs, which is the mean of the local efficiencies of all nodes: 
\begin{equation}
E_{local}=\frac{1}{n} \sum _{i \in N} E_{i, local}=\frac{1}{n} \sum _{i \in N} \frac{ \sum _{p \neq q \in N_{i}} d_{pq}^{-1}}{K_{i} \left( K_{i}-1 \right) },
\end{equation}
where $n$ is the number of nodes in the graph, $N$ is the set of all nodes, $N_{i}$ is the subgraph of neighbours of node $i$, $K_{i}$ is the degree of node $i$, and $d_{pq}$ is the shortest path length between nodes $p$ and $q$. Similar to the clustering coefficient of a graph, the local efficiency of a graph measures network segregation. As each node is not included in its own subgraph, the local efficiency of a graph reveals the fault tolerance of the network by showing how efficiently communication within the network is maintained when a node is removed, on average.

\subsubsection{Betweenness}
\paragraph{Nodal Betweenness}
Betweenness is a measure of centrality, similar to nodal degree and cost, that captures the importance of nodes in controlling information flow. It quantifies the proportion of all shortest paths in a graph that pass through a given node: 
\begin{equation}
B_{i}=\frac{1}{ \left( n-1 \right)  \left( n-2 \right) } \sum _{i \neq p \neq q \in N} \frac{ \rho _{pq}^{ \left( i \right) }}{ \rho _{pq}},
\end{equation}
where $B_{i}$ is the betweenness of node $i$, $n$ is the number of nodes in the graph, $\rho_{pq}^{ \left( i \right) }$ represents the number of shortest paths between nodes $p$ and $q$ that pass through node $i$, and $\rho_{pq}$ refers to the total number of shortest paths between nodes $p$ and $q$.
Nodal betweenness is an important metric for identifying connector hubs in a network. Nodes with high betweenness centrality can be thought of as the busiest ports in real-world networks. Removing such nodes can significantly decrease the global efficiency of the network by breaking down many short paths.

\paragraph{Mean Betweenness}
Mean betweenness refers to the betweenness of a graph. It is defined as the average nodal betweenness: 
\begin{equation}
B=\frac{1}{n} \sum _{i \in N}B_{i}=\frac{1}{n \left( n-1 \right)  \left( n-2 \right) } \sum _{i \neq p \neq q \in N} \frac{ \rho _{pq}^{ \left( i \right) }}{ \rho _{pq}}.
\end{equation}

\subsection{Statistical Analysis}
Initially, six nodal metrics, comprising degree (or cost), shortest path length, clustering coefficient, global efficiency, local efficiency, and betweenness, are computed for each region of interest (ROI) delineated in a brain atlas. Subsequently, two-tailed two-sample Student's t-tests are utilized to compare each nodal metric between two groups, namely AD and NC. Finally, the False Discovery Rate (FDR) correction is executed to rectify for numerous comparisons across 132 ROIs. Outcomes with FDR-corrected p-value less than 0.05 are deemed statistically significant.

\subsection{Results}
\subsubsection{Quality Assessment for Image Processing}
The initial phase in deriving dependable conclusions is rational and high-quality image processing. Quality evaluation and control of image processing are therefore essential. The quality assessment encompasses the manual, semi-automatic, or automatic inspection of the quality of preprocessed images. Quality control, on the other hand, refers to the actions taken in response to quality assessment. In certain situations, researchers may need to modify parameters and rerun specific experiments to obtain better image processing outcomes. In the most extreme cases, the exclusion of subjects from subsequent analyses may be necessary.
	
\begin{figure}[]
	\centering
	\includegraphics[width=0.7\linewidth]{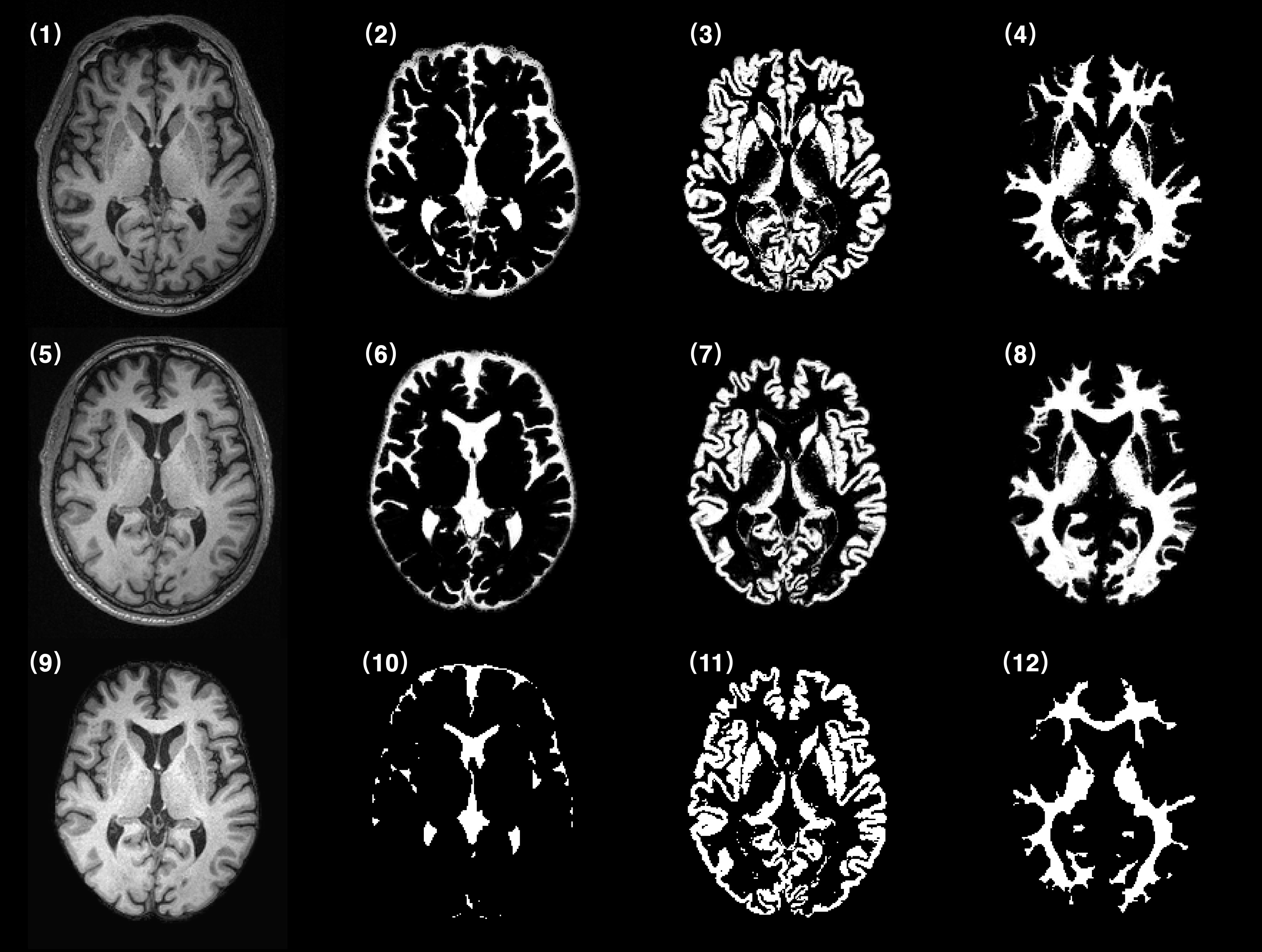}
	\caption[Segmentation of T1-weighted MRI] {{\bf Segmentation of T1-weighted MRI.} (1) Raw T1-weighted MR image in native space. (2-4) CSF, GM, WM probability maps in native space. (5) Normalized T1-weighted image in MNI-152 space. (6-8) Normalized CSF, GM, WM probability maps in MNI-152 space. (9) Skull-stripped normalized T1 image. (10-12) Normalized and eroded CSF, GM, WM masks.}
	\label{fig:segmentation.t1}
\end{figure}

\paragraph{Segmentation}
Figure \ref{fig:segmentation.t1} presents a T1-weighted MR image segmentation illustration. Precise brain tissue probability maps (sub-figures 2-4 and 6-8 in Figure \ref{fig:segmentation.t1}) can be produced through this segmentation process. The combination of segmentation and normalization also generates tissue masks in standard space (sub-figures 10-12 in Figure \ref{fig:segmentation.t1}).

\begin{figure}[]
	\centering
	\includegraphics[width=0.4\linewidth] {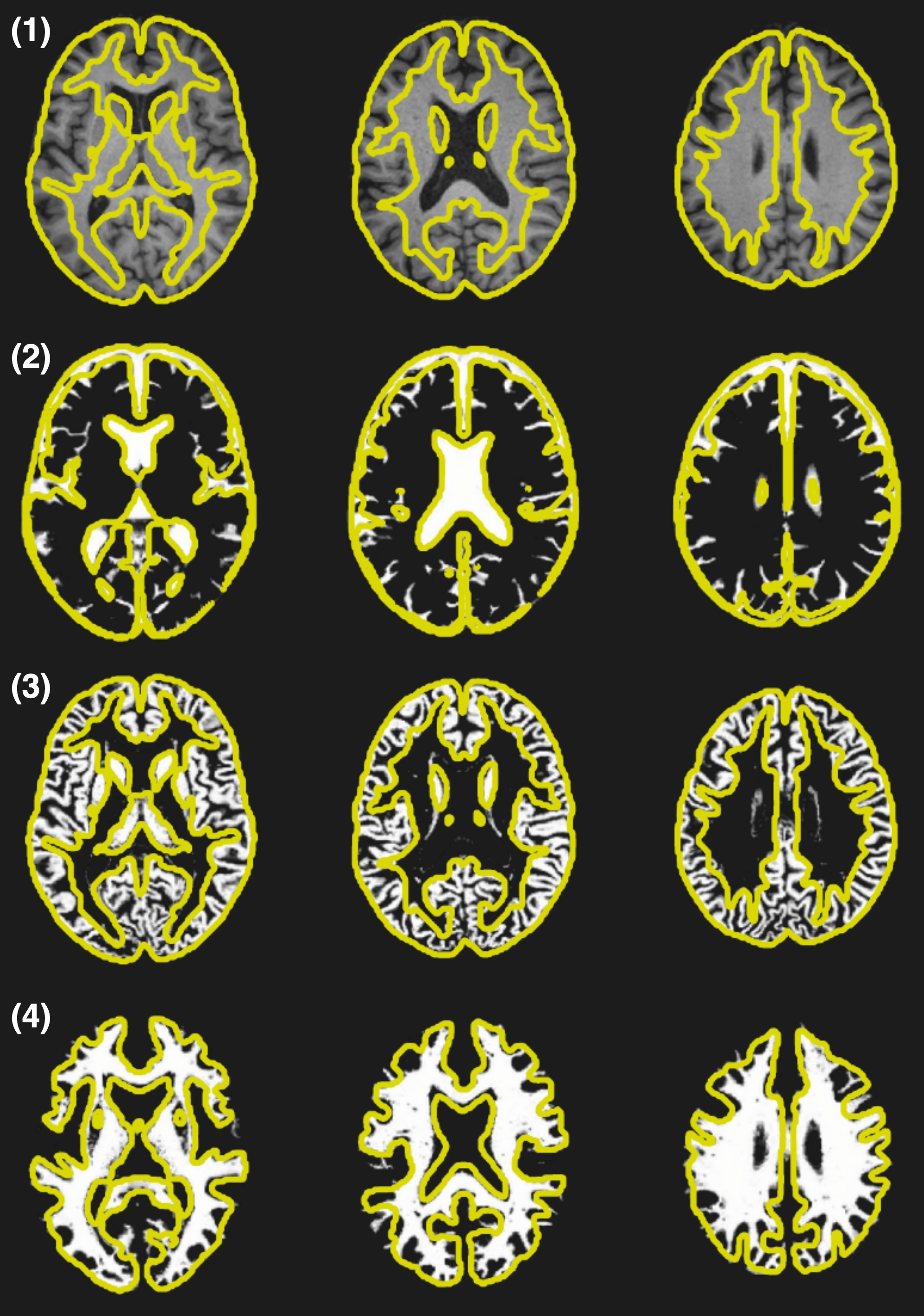}
	\caption[Normalization of T1-weighted MRI] {{\bf Normalization of T1-weighted MRI.} (1) The normalized T1-weighted MR image. (2) The normalized CSF probability map. (3) The normalized GM probability map. (4) The normalized WM probability map. All are overlapped with the contour (in yellow) of template.}
	\label{fig:normalization.t1}
\end{figure}

\begin{figure}[]
	\centering
	\includegraphics[width=0.4\linewidth] {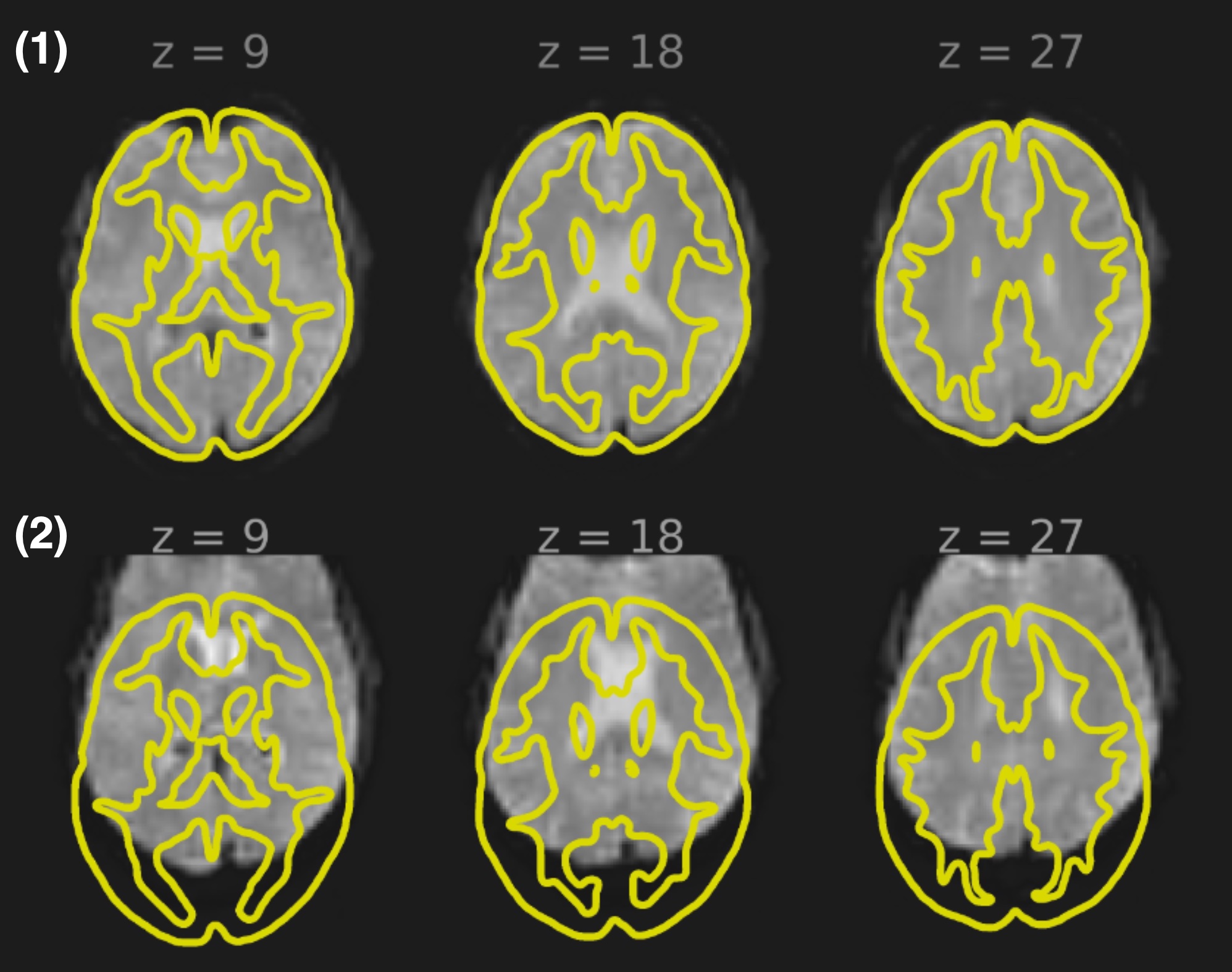}
	\caption[Normalization of BOLD fMRI] {{\bf Normalization of BOLD fMRI.} (1) The normalized BOLD fMRI overlapped with the contour (in yellow) of template. (2) The BOLD fMRI in native space overlapped with the contour (in yellow) of template.}
	\label{fig:normalization.bold}
\end{figure}

\paragraph{Normalization}
The assessment of normalization quality can be accomplished through visual inspection, wherein the contour of the template is superimposed on the normalized image. Figure \ref{fig:normalization.t1} displays the normalized T1-weighted image, CSF probability map, GM probability map, and WM probability map, which closely align with the template contour. Figure \ref{fig:normalization.bold} presents the BOLD fMRI in both standard space (Figure \ref{fig:normalization.bold} (1)) and native space (Figure \ref{fig:normalization.bold} (2)). It demonstrates that direct normalization of the BOLD MR image, which has relatively low spatial resolution, is satisfactory.

\subsubsection{Artefacts Removal}
\begin{figure}[]
	\centering
	\includegraphics[width=0.7\linewidth] {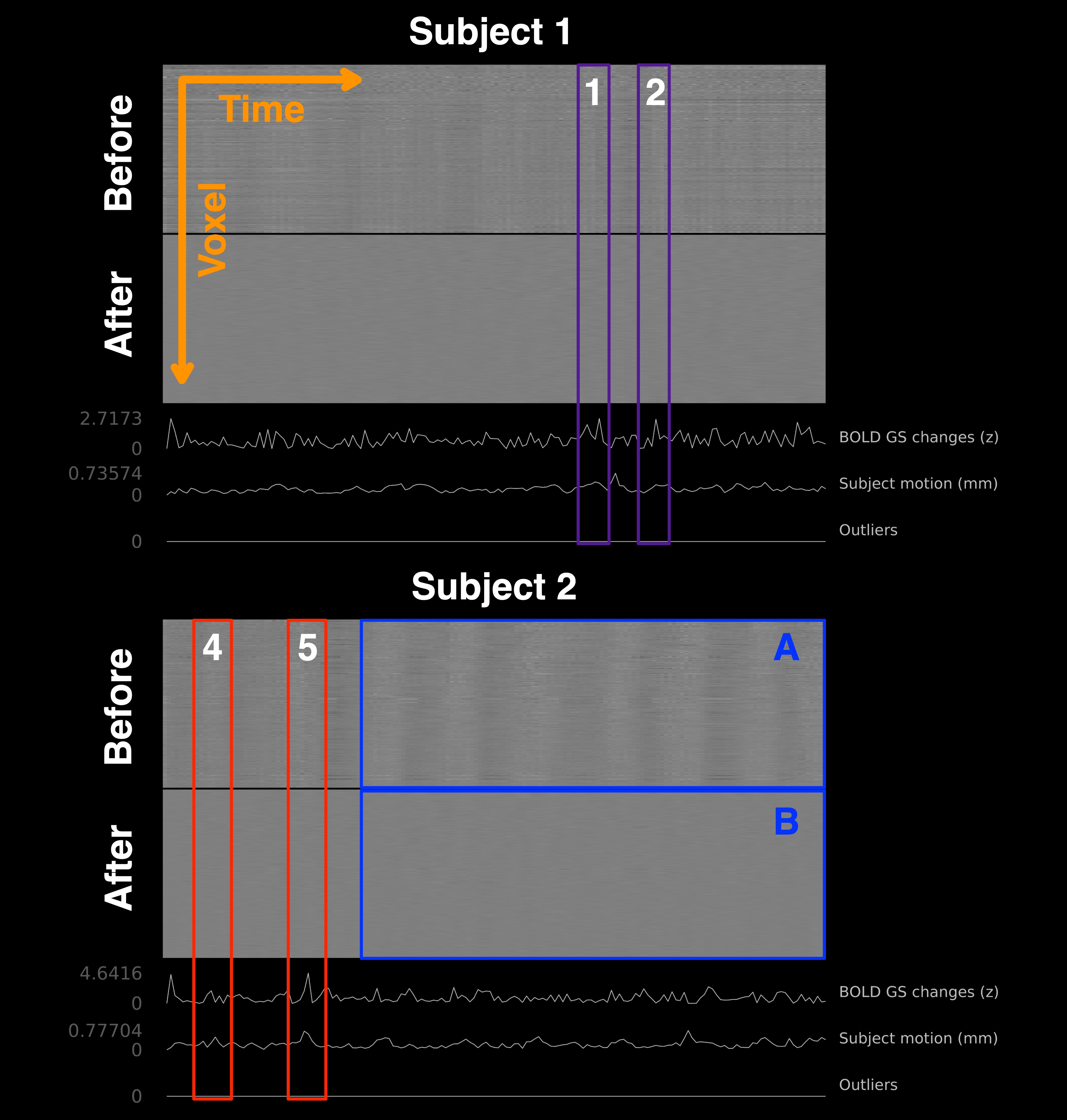}
	\caption[BOLD Signal Before and After Artefacts Removal (1)] {{\bf BOLD Signal Before and After Artefacts Removal (1).} The BOLD signals (before and after artefacts removal) of 2 subjects are shown in time series  boxes, along with which are the BOLD global signal (GS) changes in z-values, the head motion estimator, and the scrubbing variable (``outlier’’). Blue boxes (boxes A \& B) highlights the removal of global signal changes shown as  stripes pattern. Purple boxes (boxes 1 \& 2) point out the appearances of global signal changes that maybe originate from sources other than head movement. Red boxes (boxes 4 \& 5) indicate the occurrences of global signal changes that are potentially related to the subject motion.}
	\label{fig:denoising.bold_1}
\end{figure}

\begin{figure}[]
	\centering
	\includegraphics[width=0.7\linewidth] {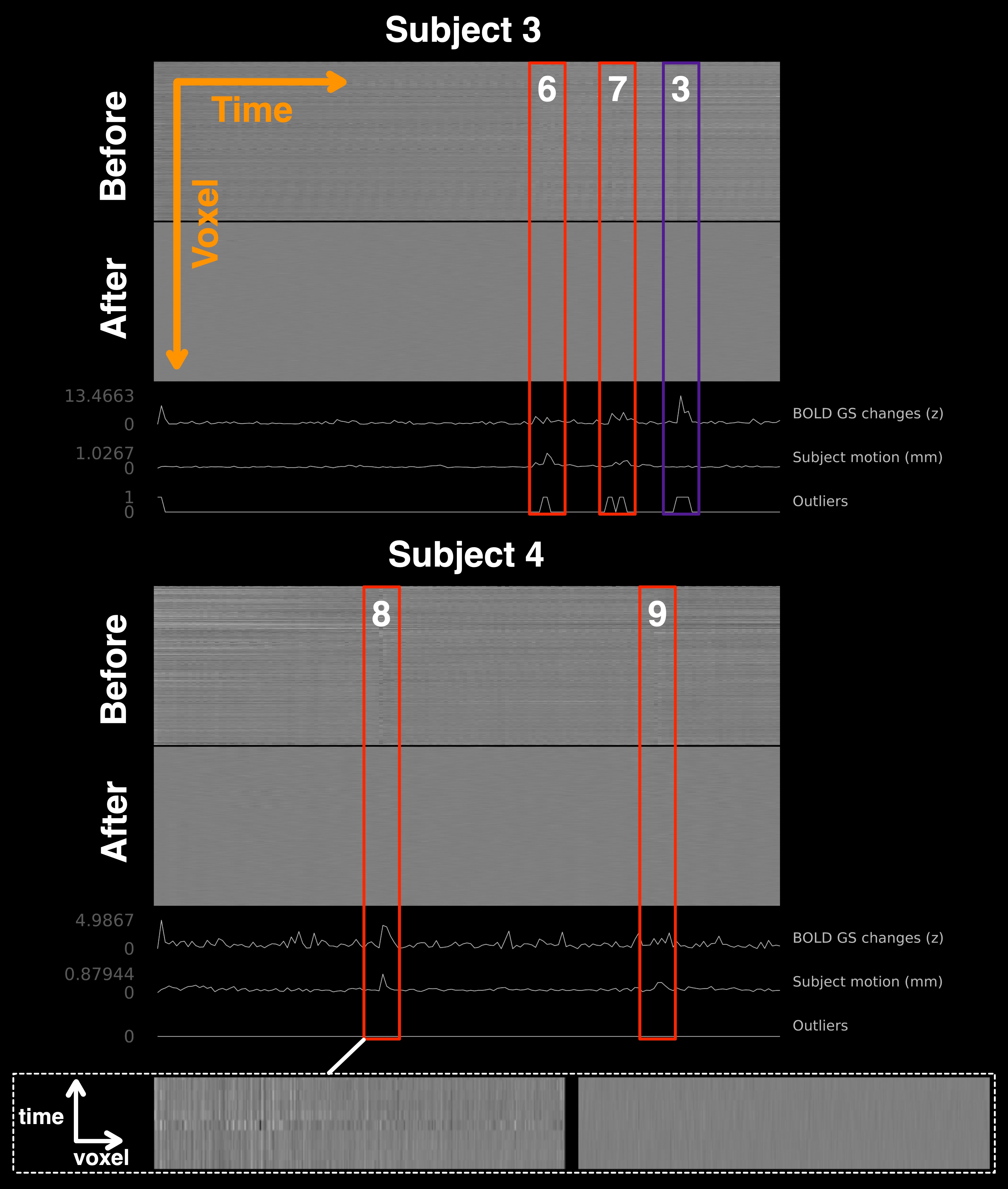}
	\caption[BOLD Signal Before and After Artefacts Removal (2)] {{\bf BOLD Signal Before and After Artefacts Removal (2).} The BOLD signals (before and after artefacts removal) of another 2 subjects are shown in time series  boxes, along with which are the BOLD global signal (GS) changes in z-values, the head motion estimator, and the scrubbing variable (``outlier’’). Purple boxes (box 3) point out the appearances of global signal changes that maybe originate from sources other than head movement. Red boxes (box 6-9) indicate the occurrences of global signal changes that are potentially related to the subject motion.}
	\label{fig:denoising.bold_2}
\end{figure}

Figure \ref{fig:denoising.bold_1} and Figure \ref{fig:denoising.bold_2} depict the BOLD signals before and after the removal of artefacts, respectively. The BOLD signals of all voxels within the GM mask are stacked vertically in time series boxes. It is evident that certain global signal variances (e.g., the striped pattern in blue box A in Figure \ref{fig:denoising.bold_1}) have been eliminated (blue box B in Figure \ref{fig:denoising.bold_1}). Alongside the BOLD signal visualization, three variables are plotted: (1) the global BOLD signal changes in z-values, (2) the integrated head motion estimator, calculated from 6 movement parameters, and (3) the scrubbing variable exported from the Artifact Detection Toolbox (ART). For the global signal changes, the mean BOLD signal (the global mean) across all GM voxels is first computed and then converted to z-values. The integrated head motion estimator from ART is a composite motion measure that estimates the maximum voxel displacement resulting from the combined effect of translation and rotation displacement measures. The scrubbing variable is a frame identifier pointing out the outlier scans.

Note that two types of global signal variances are highlighted in Figure \ref{fig:denoising.bold_1} and Figure \ref{fig:denoising.bold_2}. In each red box, we can see that sudden global signal changes always occur with severe head motion, particularly in box 6 and 7 (Figure \ref{fig:denoising.bold_2}), where MR scans are identified as outliers at the same time point. Therefore, we can conclude that the red boxes (boxes 4 and 5 in Figure \ref{fig:denoising.bold_1}, boxes 6-9 in Figure \ref{fig:denoising.bold_2}) indicate global signal changes potentially related to subject motion. The purple boxes (boxes 1 and 2 in Figure \ref{fig:denoising.bold_1}, box 3 in Figure \ref{fig:denoising.bold_2}) highlight global signal changes that may originate from sources other than head movement.

\begin{figure}[]
	\centering
	\includegraphics[width=0.5\linewidth] {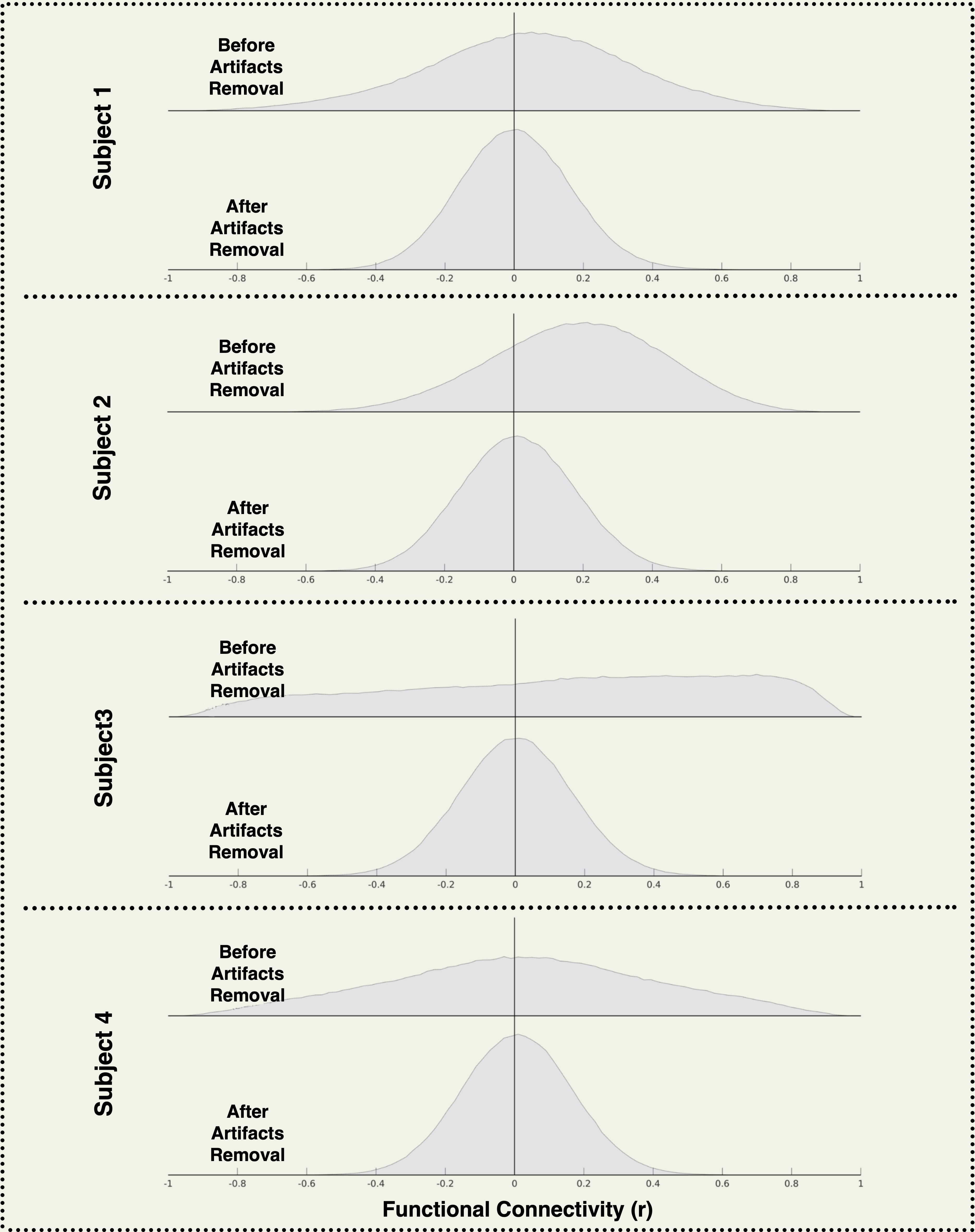}
	\caption[Distribution of FCs Before and After Artefacts Removal] {{\bf Distribution of FCs Before and After Artefacts Removal.} The distribution of functional connections (FCs) of 4 subjects are shown. Before artefacts removal, the distributions are right-skewed and sometimes non-bell-shaped. After artefacts removal, the FCs distributions become normal with approximate zero-mean and bell-shape.}
	\label{fig:denoising.fc}
\end{figure}

The impact of artefacts removal on the distribution of functional connections (FCs) can be observed from Figure \ref{fig:denoising.fc}. First, GM voxels are segmented into 1000 clusters, defining 1000 nodes. Second, mean BOLD signals are computed from each cluster. Third, Pearson's linear correlation coefficient ($r$) is used to measure FC. Finally, the distribution of FCs is plotted. Similar effects can be observed in four subjects. The FCs distribution before artefacts removal exhibits varying degrees of positive skewness, indicating the presence of a systematic bias towards positive FCs. However, after removing artefacts, it becomes approximately normally distributed with a mean close to zero.

\begin{figure}[]
	\centering
	\includegraphics[width=0.5\linewidth] {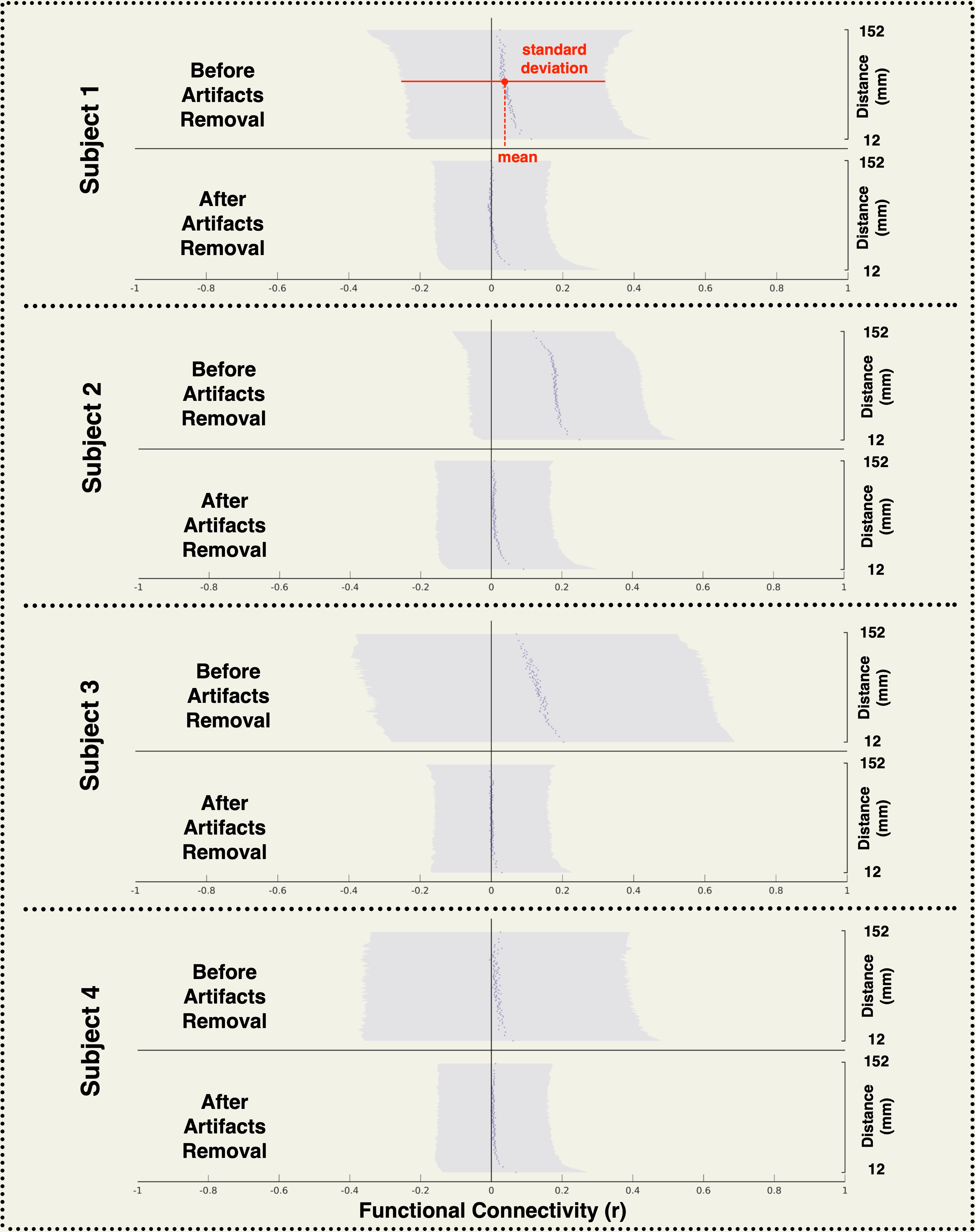}
	\caption[Distance-FCs Map Before and After Artefacts Removal] {{\bf Distance-FCs Map Before and After Artefacts Removal.} The figure shows the functional connectivity plotted against distance (in $mm$). Each scattering dot represents the mean functional connection (FC) at a specific distance, and the shadow area indicates the standard deviation. Similar to the findings in Figure \ref{fig:denoising.fc}, FCs are shifted to the positive side before artefacts removal. After artefacts removal, most of the mean FCs are near zero except the short-distance ones.}
	\label{fig:denoising.fc.distance}
\end{figure}

\begin{figure}[]
	\centering
	\includegraphics[width=0.5\linewidth] {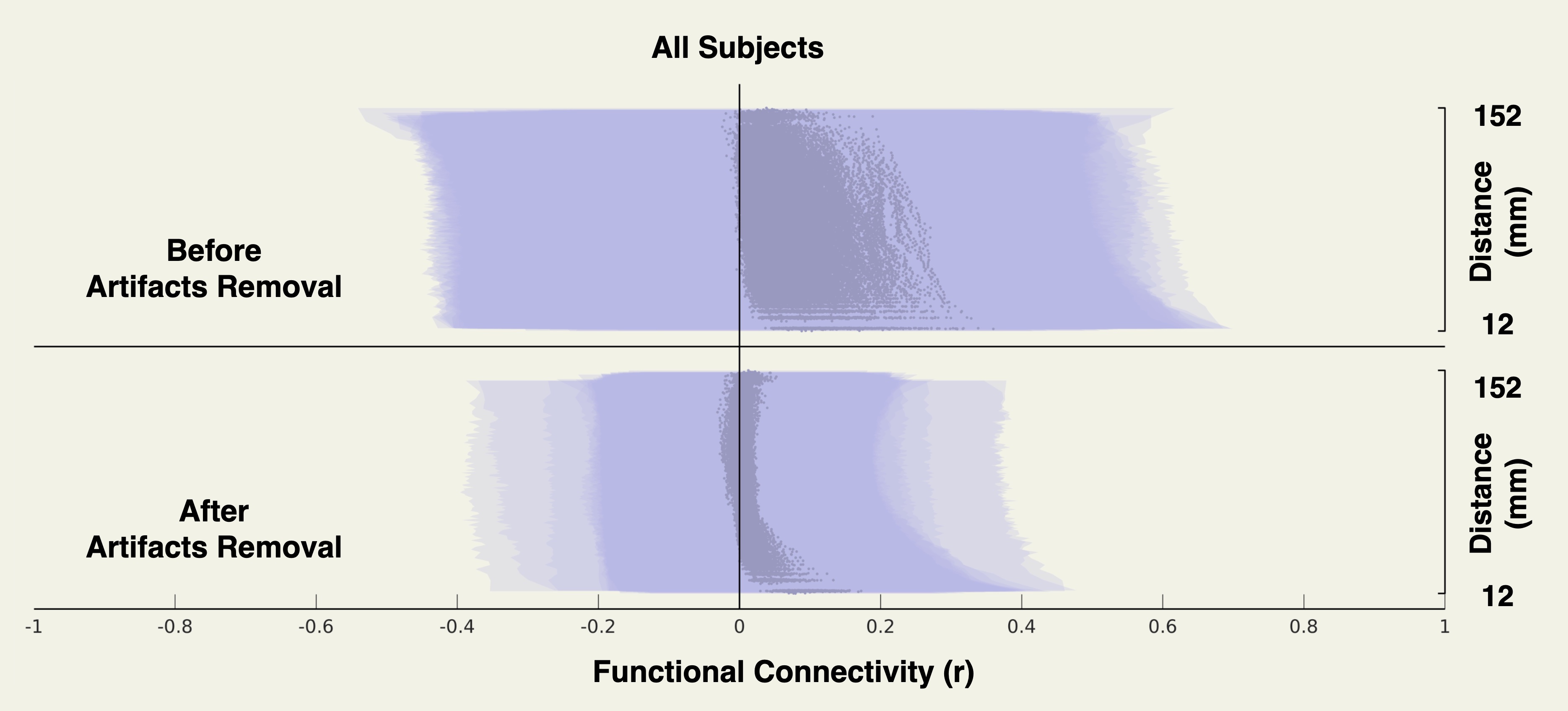}
	\caption[Distance-FCs Map Before and After Artefacts Removal (All Subjects)] {{\bf Distance-FCs Map Before and After Artefacts Removal (All Subjects).} The distributions and distance-related properties of FCs before and after artefacts removal are shown. It suggests that FCs are less divergent and more zero-centred in mediate to long distances after artefacts removal. Short-distance FCs are stronger compared with their long-distance counterparts.}
	\label{fig:denoising.fc.distance.allsub}
\end{figure}

Furthermore, Figure \ref{fig:denoising.fc.distance} provides additional information on FCs by incorporating distance-related data. The definition of FCs in Figure \ref{fig:denoising.fc.distance} is the same as in Figure \ref{fig:denoising.fc}. To draw the distance-FCs maps in Figure \ref{fig:denoising.fc.distance}, an extra step is required: calculating the distance (in $mm$) between two nodes. As shown in the first row of Figure \ref{fig:denoising.fc.distance}, each scattered dot represents the mean FC at a specific distance, and the shaded area indicates the standard deviation. Compared to the FCs distributions in Figure \ref{fig:denoising.fc}, the FCs are shifted towards positive values before removing artefacts. After artefacts removal, most of the mean FCs are near zero, except for short-distance FCs.
The distance-FCs maps for four subjects (Figure \ref{fig:denoising.fc.distance}) demonstrate similar distance-related FCs properties before and after artefacts removal. They also indicate that removing artefacts can reduce the bias towards positive values in mediate and long distance FCs. This effect is confirmed in Figure \ref{fig:denoising.fc.distance.allsub}, where the distance-FCs maps for all subjects are superimposed on top of each other. It shows that, after removing artefacts, FCs are less divergent and more centred around zero at mediate to long distances, while short-distance FCs are stronger than their mediate and long-distance counterparts.

\subsection{Abnormal Functional Brain Network Properties in AD}
As described in the method section, three types of networks are constructed to explore the effects of negative functional connections (anti-correlations). The positive network is created by retaining only the top $15\%$ positive FCs. The negative network is constructed by retaining the top $15\%$ negative FCs. In the mixed network, the anti-correlations are first replaced by their absolute values and then thresholded in the same manner as the positive network.

\subsubsection{Abnormalities in Positive Network}
Table \ref{tab:result.graph.posi.metrics} illustrates significant between-group differences (FDR-p $<$ $0.05$) found in six nodal metrics, namely local efficiency, clustering coefficient, degree, cost, average shortest path length, and global efficiency. The results consistently indicate a singular direction of abnormality. A summary of the statistically significant findings is presented below.

\begin{small}
\begin{table}[h]
	\centering
	\caption{Abnormal Graph Metrics in AD (Results from Positive Network)}
	\label{tab:result.graph.posi.metrics}
\begin{tabular}{l | lccc}
\hline \hline
\multicolumn{1}{c}{\textbf{Metrics}} & \multicolumn{1}{c}{\textbf{Region}} & \makecell{\textbf{T Statistics} \\ \textbf{(AD - NC)}} & {\bf p-value} & {\bf FDR-p} \bigstrut \\
\hline \hline
\multirow{8}[0]{*}{Local Efficiency} & PT l  & -4.59 & 0.000005 & 0.000701 \bigstrut[t] \\
      & pMTG r & -3.70 & 0.000234 & 0.014016 \\
      & PT r  & -3.62 & 0.000319 & 0.014016 \\
      & CO r  & -3.28 & 0.001113 & 0.036481 \\
      & aSTG r & -3.21 & 0.001382 & 0.036481 \\
      & HG r  & -3.12 & 0.001882 & 0.041105 \\
      & Cuneal l & -3.08 & 0.002180 & 0.041105 \\
      & CO l  & -2.99 & 0.002913 & 0.048057 \bigstrut[b] \\
\hline
Clustering Coefficient & PT l  & -4.26 & 0.000024 & 0.003168 \bigstrut \\
\hline
\multirow{3}[0]{*}{\makecell{Average Shortest \\ Path Length}} & Cereb3 r & -3.96 & 0.000083 & 0.010937 \bigstrut[t] \\
      & Ver8  & -3.69 & 0.000246 & 0.012355 \\
      & Cereb3 l & -3.65 & 0.000281 & 0.012355 \bigstrut[b] \\
\hline
\multirow{8}[0]{*}{Global Efficiency} & Cereb3 r & 4.53  & 0.000007 & 0.000948 \bigstrut[t] \\
      & Cereb3 l & 3.64  & 0.000293 & 0.018702 \\
      & Cereb6 r & 3.54  & 0.000425 & 0.018702 \\
      & Cereb45 r & 3.32  & 0.000965 & 0.026026 \\
      & Ver45 & 3.31  & 0.000986 & 0.026026 \\
      & Ver3  & 3.24  & 0.001274 & 0.028018 \\
      & Ver8  & 3.12  & 0.001900 & 0.035825 \\
      & Cereb8 r & 3.06  & 0.002285 & 0.037695 \bigstrut[b] \\
\hline
\multirow{2}[0]{*}{Degree / Cost} & Cereb3 r & 4.39  & 0.000013 & 0.001744 \bigstrut[t] \\
      & Cereb6 r & 3.47  & 0.000550 & 0.036294 \bigstrut[b] \\
\hline
\hline
\end{tabular}
\end{table}
\end{small}

\begin{figure}[]
	\centering
	\includegraphics[width=0.8\linewidth] {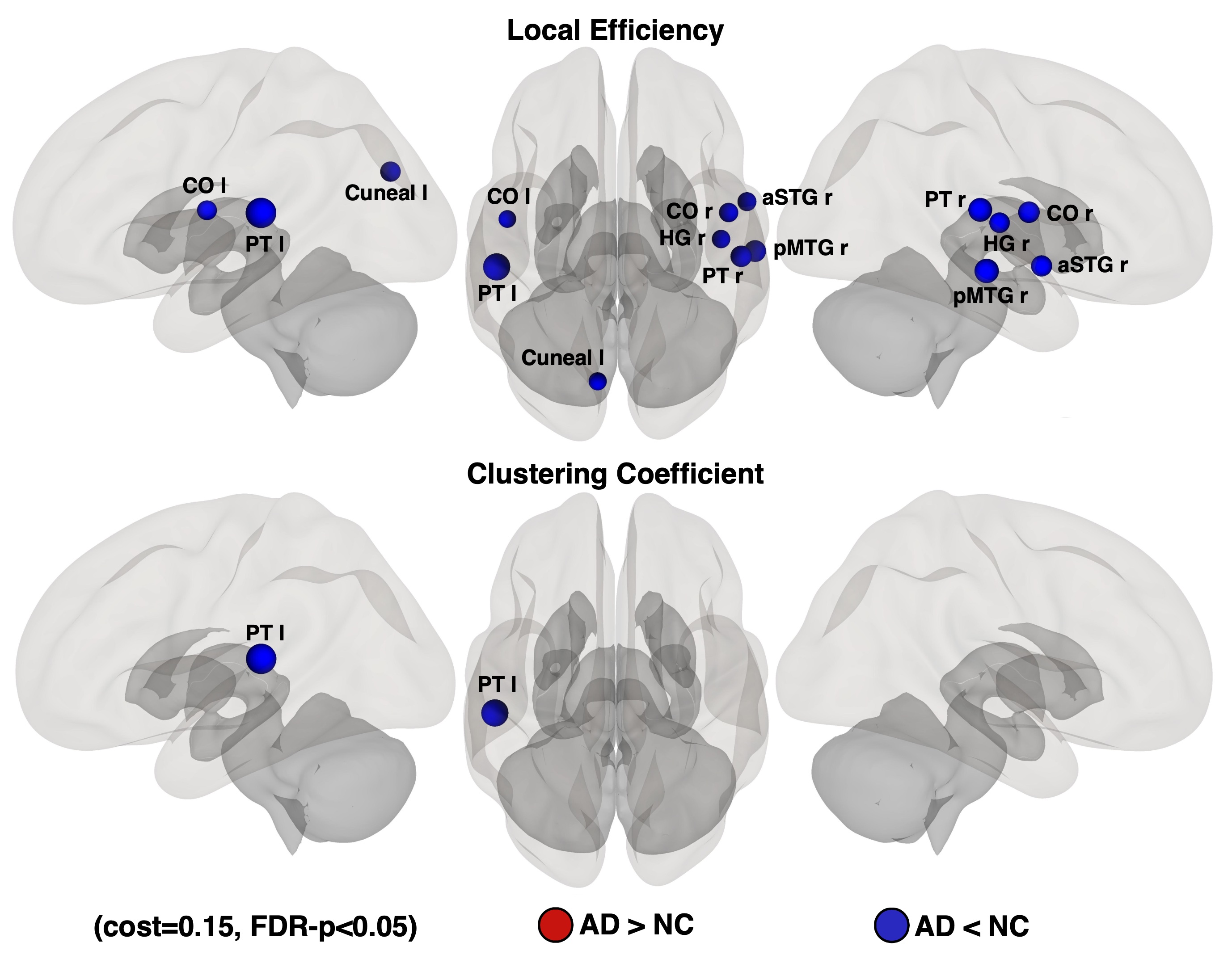}
	\caption[Abnormal Graph Metrics in Cortical Cortices and Subcortical Regions (Results from Positive Network)] {{\bf Abnormal Graph Metrics in Cortical Cortices and Subcortical Regions (Results from Positive Network).} l: left; r: right; aSTG: anterior Superior Temporal Gyrus; PT: Planum Temporale; pMTG: posterior Middle Temporal Gyrus; CO: Central Opercular Cortex; Cuneal: Cuneal Cortex; HG: Heschl’s Gyrus.}
	\label{fig:result.graph.posi.net.cortical}
\end{figure}

\begin{figure}[]
	\centering
	\includegraphics[width=0.8\linewidth] {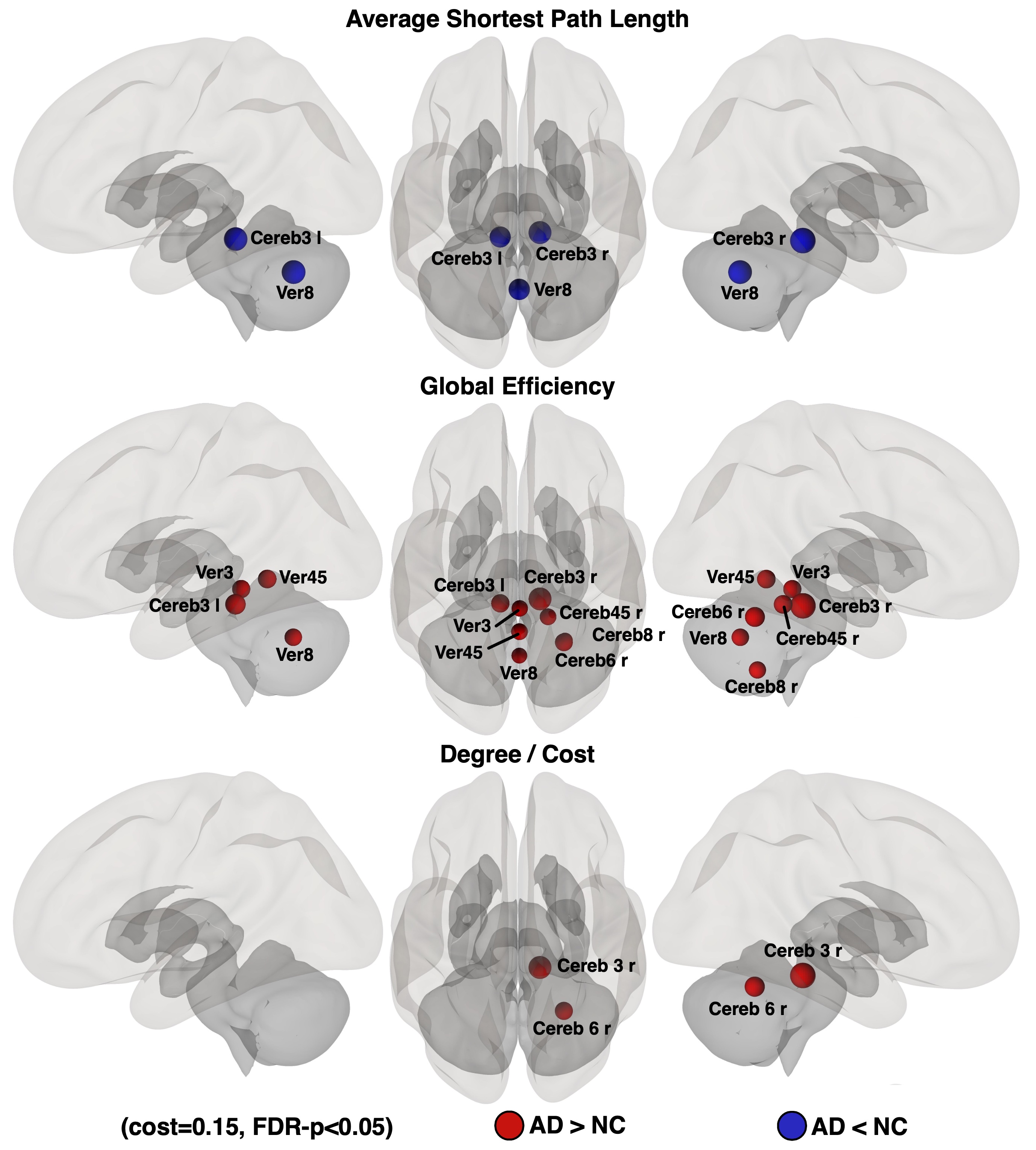}
	\caption[Abnormal Graph Metrics in Cerebellum (Results from Positive Network)] {{\bf Abnormal Graph Metrics in Cerebellum (Results from Positive Network).} l: left; r: right; Cereb: Cerebellum; Ver: Vermis.}
	\label{fig:result.graph.posi.net.cereb}
\end{figure}

\begin{itemize}
\item Compared with the NC group, decreased local efficiencies are discovered in the bilateral Central Opercular Cortex (CO), the bilateral Planum Temporale (PT), the left Cuneal, the right Heschl’s Gyrus (HG), the right posterior Middle Temporal Gyrus (pMTG), and the right anterior Superior Temporal Gyrus (aSTG) (Figure \ref{fig:result.graph.posi.net.cortical}) in AD group.

\item Compared with the NC group, the decreased clustering coefficient is observed in the left Planum Temporale (PT) (Figure \ref{fig:result.graph.posi.net.cortical})  in the AD group.

\item Compared with the NC group, increased degrees and costs are only found in the cerebellum, including the area 3 and area 6 of the right Cerebellum (Cereb3 \& Cereb6) (Figure \ref{fig:result.graph.posi.net.cereb}) in the AD group.

\item Compared with the NC group, decreased average shortest path lengths are observed in area 3 of the bilateral Cerebellum (Cereb3) and area 8 of the Vermis (Ver 8) (Figure \ref{fig:result.graph.posi.net.cereb}) in the AD group.

\item Compared with the NC group, increased global efficiencies are solely found in the Cerebellum, including area 3 of the bilateral Cerebellum (Cereb3), the area 4 \& 5, area 6 and area 8 of the right Cerebellum (Cereb45, Cereb6 and Cereb8), and the area 3, 4 \& 5, and 8 of the Vermis (Ver3, Ver45, and Ver8) (Figure \ref{fig:result.graph.posi.net.cereb}) in the AD group.

\end{itemize}
	
\subsubsection{Abnormalities in Mixed Network}
As shown in Table \ref{tab:result.graph.mixed.metrics}, similar but less group-differences (FDR-p $<$ $0.05$) are observed. Below is a summary of the results.

\begin{small}
\begin{table}[h]
	\centering
	\caption{Abnormal Graph Metrics in AD (Results from Mixed Network)}
	\label{tab:result.graph.mixed.metrics}
\begin{tabular}{l | lccc}
\hline \hline
\multicolumn{1}{c}{\textbf{Metrics}} & \multicolumn{1}{c}{\textbf{Region}} & \makecell{\textbf{T Statistics} \\ \textbf{(AD - NC)}} & {\bf p-value} & {\bf FDR-p} \bigstrut \\
\hline \hline
\multirow{5}[2]{*}{Local Efficiency} & pMTG r & -4.88 & 0.000001 & 0.000179 \bigstrut[t]\\
      & CO r  & -4.32 & 0.000018 & 0.001188 \\
      & PT l  & -3.68 & 0.000250 & 0.010996 \\
      & aSTG l & -3.49 & 0.000519 & 0.017134 \\
      & pMTG l & -3.15 & 0.001714 & 0.045251 \bigstrut[b]\\
\hline
\multirow{3}[2]{*}{Clustering Coefficient} & pMTG r & -3.97 & 0.000081 & 0.010721 \bigstrut[t]\\
      & CO r  & -3.67 & 0.000265 & 0.017475 \\
      & PT l  & -3.45 & 0.000601 & 0.026451 \bigstrut[b]\\
\hline
Global Efficiency & Cereb3 l & 3.69  & 0.000242 & 0.031955 \bigstrut\\
\hline
\hline
\end{tabular}%
\end{table}
\end{small}

\begin{figure}[]
	\centering
	\includegraphics[width=0.8\linewidth] {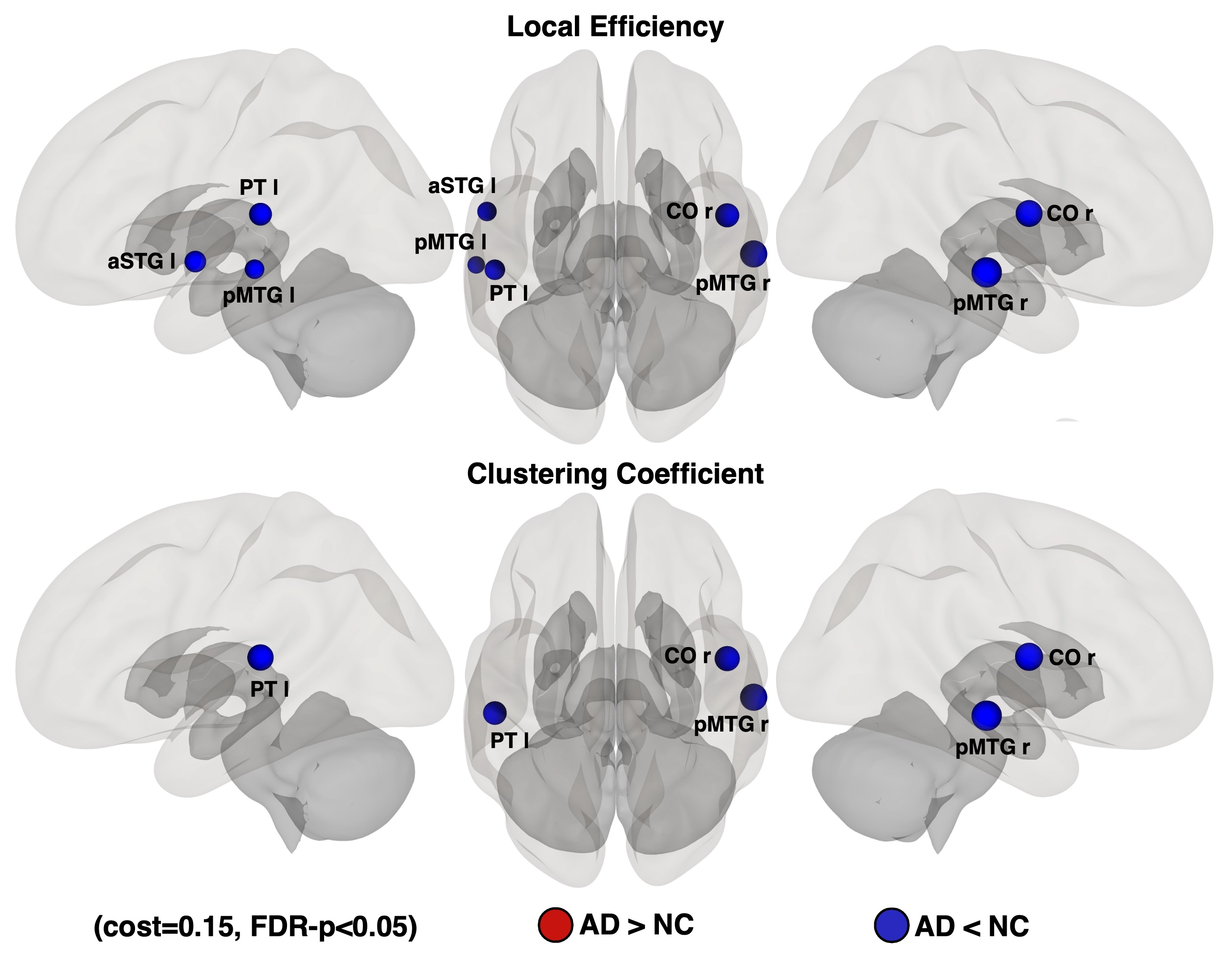}
	\caption[Abnormal Graph Metrics in Cortical Cortices and Subcortical Regions (Results from Mixed Network)] {{\bf Abnormal Graph Metrics in Cortical Cortices and Subcortical Regions (Results from Mixed Network).} l: left; r: right; aSTG: anterior Superior Temporal Gyrus; PT: Planum Temporale; pMTG: posterior Middle Temporal Gyrus; CO: Central Opercular Cortex.}
	\label{fig:result.graph.mixed.net.cortical}
\end{figure}

\begin{figure}[]
	\centering
	\includegraphics[width=0.8\linewidth] {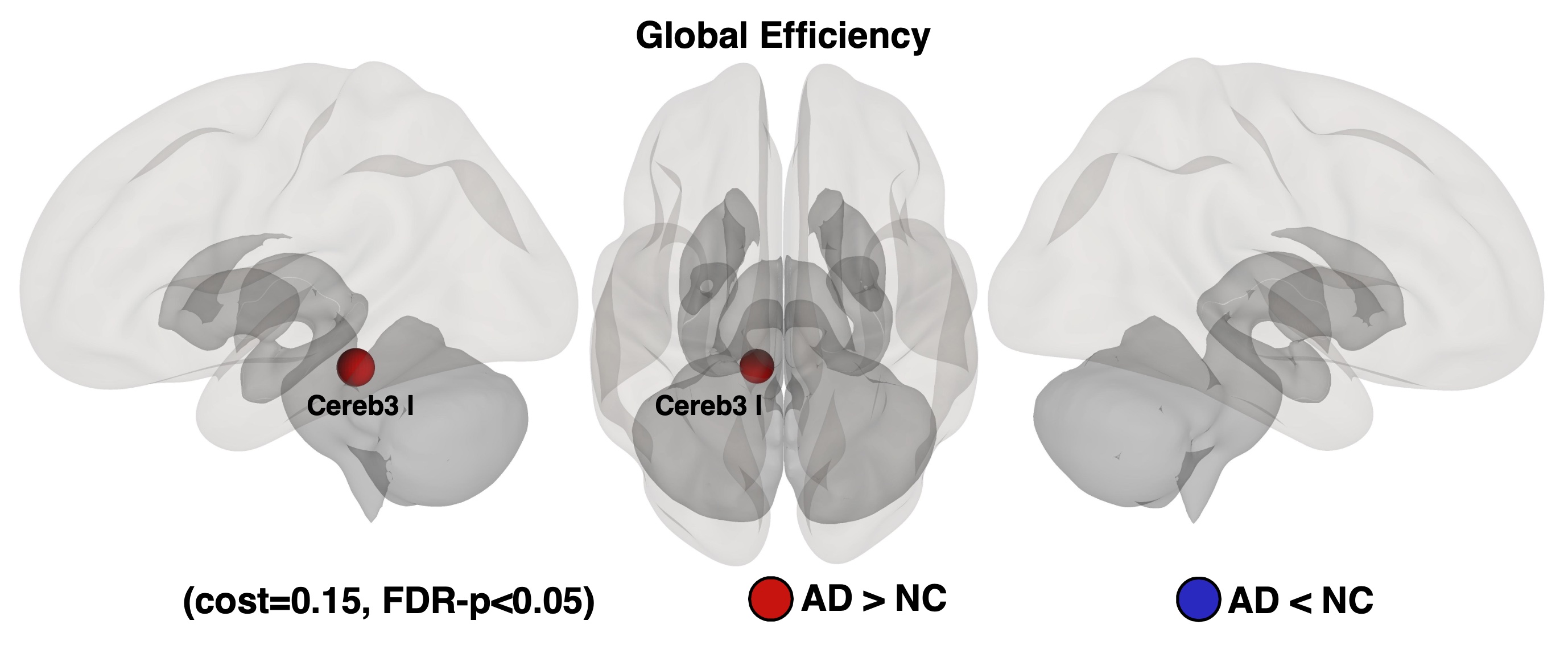}
	\caption[Abnormal Graph Metrics in Cerebellum (Results from Mixed Network)] {{\bf Abnormal Graph Metrics in Cerebellum (Results from Mixed Network).} l: left; r: right; Cereb: Cerebellum.}
	\label{fig:result.graph.mixed.net.cereb}
\end{figure}

\begin{itemize}
\item Compared with the NC group, decreased local efficiencies are discovered in the AD group in the bilateral posterior Middle Temporal Gyrus (pMTG), the left Planum Temporale (PT), the left anterior Superior Temporal Gyrus (aSTG), and the right Central Opercular Cortex (CO) (Figure \ref{fig:result.graph.mixed.net.cortical}).

\item Compared with the NC group, decreased clustering coefficients are observed in the AD group in the left Planum Temporale (PT), the right Central Opercular Cortex (CO), and the right posterior Middle Temporal Gyrus (pMTG) (Figure \ref{fig:result.graph.mixed.net.cortical}).

\item Compared with the NC group, increased global efficiency is only found in the AD group in area 3 of the left Cerebellum (Cereb3) (Figure \ref{fig:result.graph.mixed.net.cereb}).
\end{itemize}

\subsection{Discussion}
\subsubsection{Removal of Global Effects}
\paragraph{Global Signal}
In functional neuroimaging, the global signal is defined as the mean time series averaged across all brain voxels, including those in gray matter (GM), white matter (WM), and cerebrospinal fluid (CSF). GM is typically the region of interest (ROI) in brain fMRI studies because the dense neurons allow for indirect reflection of neural activity through fluctuations in the BOLD signal. However, the BOLD contrast is an indirect measure resulting from complex interactions between mainly three factors: cerebral metabolic rate of oxygen (CMRO2), blood volume, and blood flow. Thus, any factor that disrupts the balance between these three parameters can alter the BOLD signal. For example, head motion \cite{power2014methods}, cardiac and respiratory cycles \cite{chang2009effects}, arterial CO2 concentration, cerebral autoregulation, blood pressure, and vasomotion \cite{murphy2013resting} can introduce non-neuronal variances to all BOLD time series. Consequently, the global effects of artefacts can cause two BOLD signals to be more statistically correlated, and the distribution of correlations between BOLD signals is heavily skewed towards positive values. Many non-neuronal confounds can contribute to the global signal. However, it has also been demonstrated that the global signal can be tightly linked to neural activity \cite{scholvinck2010neural}.

\paragraph{Global Signal Regression}
Global Signal Regression (GSR) is a method proposed to remove the confounding effects of non-neuronal sources \cite{macey2004method,aguirre1998inferential}. GSR employs linear regression for each voxel to remove variances explained by the global signal, and the resulting residuals can be considered as the de-noised signal. One of the most significant and widely discussed effects of GSR is the emergence of more anti-correlations. The mathematical details of the algebraic operation of GSR and the impact of mandating anti-correlations have been published \cite{murphy2009impact,fox2009global}. It is explained that mathematically, the seed-to-voxel regression coefficients ($\beta$) have a mean of exactly zero, and the distribution of Pearson's linear correlation coefficients is approximately zero-centred with GSR. Murphy and colleagues \cite{murphy2009impact} were the first to mathematically demonstrate that GSR forces approximately half of the correlations to become anti-correlations and concluded that the anti-correlations between task-negative regions (i.e., DMN regions) and task-positive regions are most likely an outcome of GSR. Conversely, other researchers \cite{fox2009global,chai2012anticorrelations} support the idea that the anti-correlations after GSR have biological origins. To date, whether GSR should be applied remains an open question, and contradictory recommendations have been made \cite{murphy2017towards}. Without a "gold standard," it is difficult to determine whether anti-correlation is merely an artefact of the pre-processing strategy or physiologically meaningful.

\paragraph{Drawbacks of GSR}
On the one hand, it has been reported that GSR can not only eliminate non-neuronal artefacts but also enhance the specificity of positive correlations and their correspondence to structural connectivity \cite{fox2009global}. On the other hand, the drawbacks of GSR have been discussed. GSR can introduce spurious negative correlations between brain regions \cite{murphy2009impact,anderson2011network,saad2012trouble} and alter the group comparison of inter-regional correlations \cite{saad2012trouble}. Murphy and colleagues \cite{murphy2009impact} used simulated data to demonstrate that GSR is not effective in removing nuisance confounds, and the locations of anti-correlated areas depend on the relative phases of the global signal and seed voxel time series. Saad and colleagues \cite{saad2012trouble} used an illustrative model to show that GSR distorts the inter-regional correlations in a way that depends on the true correlation structure of the entire network and regional size distribution. Anderson and colleagues \cite{anderson2011network} also demonstrated that anti-correlations after GSR can be introduced even in completely uncorrelated networks.

\paragraph{Alternatives to GSR}
Since GSR adds uncertainty and interpretive complexity to scientific findings, various alternatives to GSR have been explored and proposed. All of these alternatives share the same motivation as GSR, which is to remove the common non-neuronal BOLD signal fluctuations and reveal the true functional connectivity. One type of technique involves the inclusion of physiological data. Physiological signals must be recorded simultaneously with fMRI data acquisition. When these physiological data are not available, there are still methods to remove global effects. For example, the aCompCor technique \cite{behzadi2007component} uses anatomical masks to increase spatial specificity and principal components instead of mean signal to eliminate confounding artefacts. More complex methods, such as the Random Subspace Method for Functional Connectivity (RSMFC) and the Affine Parameterization of Physiological Large-Scale Error Correction (APPLECOOR), are summarized in the review paper by Kevin Murphy and Michael Fox \cite{murphy2017towards}.

\paragraph{Reasons for Using aCompCor}
The current study is a functional network study based on graph theory that aims to investigate topological reconfiguration. Anti-correlations are treated differently in three network construction strategies: (1) only positive correlations are used for network construction in the analysis of positive networks; (2) similarly, only anti-correlations are used for negative networks; and (3) the absolute values of positive and negative correlations are used for mixed networks. By comparing the results of these three experiments, the effects of anti-correlations can be investigated. Furthermore, it has been reported that both local and global topological properties, quantified by local and global graph metrics, have higher reliability when GSR is not applied compared to when it is applied \cite{liang2012effects}.

\subsubsection{Comparison of Findings in Positive and Mixed Network}
	For each network metric, consistent results are observed for both positive and mixed networks. Here, the consistency is reflected not only by the same direction of alteration, but not also by the rationality of network metrics alterations. A summary of the consistency of findings from the positive and mixed network are as followed.

\paragraph{The finding are not contradictory to each others.} Results (Table \ref{tab:result.graph.posi.metrics} and Table \ref{tab:result.graph.mixed.metrics}) show that only decreased local efficiency, decreased clustering coefficient, decreased average shortest path length, increased global efficiency, and increased degree (or cost) are found.

\begin{itemize}
\item Decreased local efficiency and decreased clustering coefficient are consistent. By definition, the local efficiency of node $i$ is the global efficiency of the subgraph ($N_{i}$) of its neighbouring nodes. The clustering coefficient of node $i$ is defined as the proportion of edges (connections) among all the possible edges in the subgraph ($N_{i}$). Importantly, the global efficiency is directly and inversely related to the shortest path length: it is defined as the harmonic mean of the shortest path lengths. The logical connection between local efficiency and clustering efficient can be seen in the following example. If the neighbours of node $i$ are densely interconnected which is equivalent to a high clustering coefficient for node $i$, the shortest path length for each pair of nodes in $N_{i}$ would be low which would lead to high global efficiency for subgraph $N_{i}$. And high global efficiency for subgraph $N_{i}$ is equal to high local efficiency for node $i$, by definition. Therefore, the clustering coefficient and local efficiency are positively correlated. For example, decreased local efficiency and decreased clustering coefficient are found in left Planum Temporale (PT) (Table \ref{tab:result.graph.posi.metrics} \& Table \ref{tab:result.graph.mixed.metrics}).
\item Decreased average shortest path length and increased global efficiency are consistent. As explained above, the global efficiency of node $i$ is defined as the harmonic mean of all shortest path length between node $i$ and all other nodes in the same graph. Thus, the shortest path length and global efficiency are negatively correlated. 
\item Increased degree (or cost), increased global efficiency and decreased average shortest path length are consistent. Although degree and cost are not directly related to global efficiency in their definitions, they are more likely to be positively correlated to the average shortest path length and global efficiency. Supposed node $i$ has a high degree which means node $i$ is directly connected with many nodes, a great deal of shortest paths emitted from node $i$ will have length 1, decreasing the average shortest path length for node $i$. For example, decreased average shortest path length, increased global efficiency, and increased degree and cost are found in area 3 of the right Cerebellum (Cereb3) (Table \ref{tab:result.graph.posi.metrics}).
\item Global efficiency quantifies network integration, while local efficiency quantifies network segregation. Though the altered global efficiencies and local efficiencies found in the AD group are in opposite direction, the author holds the opinion that they depict distinct aspects of network properties, and are not strongly correlated.
\end{itemize}

\paragraph{The altered network properties found in the mixed network is consistent with that of the positive network, but less in the amount.}

\begin{itemize}
\item Decreased local efficiencies are found in the left Planum Temporale (PT), right posterior Middle Temporal Gyrus (pMTG), and right Central Opercular Cortex (CO) in both positive and mixed network (Table \ref{tab:result.graph.posi.metrics} \& Table \ref{tab:result.graph.mixed.metrics}). 
\item Decreased clustering coefficient is found in left Planum Temporale (PT) in the both positive and mixed network (Table \ref{tab:result.graph.posi.metrics} \& Table \ref{tab:result.graph.mixed.metrics}).
\item Increased global efficiency is found in left area 3 of right Cerebellum (Cereb3) in the both positive and mixed network (Table \ref{tab:result.graph.posi.metrics} \& Table \ref{tab:result.graph.mixed.metrics}).
\item Altered metrics found only in the positive network include: decreased local efficiency (in right PT, right aSTG, right HG, left Cuneal, and left CO), decreased average shortest path length (in bilateral Cereb3 and Ver8), increased global efficiency (in right Cereb3, right Cereb6, right Cereb45, Ver45, Ver3, Ver8, and right Cereb8), and increased degree and cost (in right Cereb3 and right Cereb6) (Table \ref{tab:result.graph.posi.metrics}).
\item Altered metrics found only in the mixed network include decreased local efficiency (in left aSTG and left pMTG), decreased clustering coefficient (in right pMTG and right CO) (Table \ref{tab:result.graph.mixed.metrics}).
\end{itemize}

\paragraph{Compared with NC, weaker functional segregation is discovered in the AD group in only cortical regions.} Decreased local efficiencies and clustering coefficients, which are two measures of network segregation, are only observed in cortical cortices, but not in any region of subcortical structures and cerebellum. 

\paragraph{Compared with NC, stronger functional integration is discovered in the AD group in only regions of subcortical structures and cerebellum.} Decreased average shortest path length, increased global efficiency, and increased degree, which are measures of network integration, are only observed in subcortical structures and cerebellum.

\bibliographystyle{unsrt}  
\bibliography{references}

\end{document}